\documentclass[aps,pre,showpacs,
amssymb,twocolumn
]{revtex4-2}
\usepackage{graphicx}
\usepackage{amsmath}
\usepackage{amsfonts}
\usepackage{amssymb}
\usepackage{epsfig,libertine}
\usepackage[libertine]{newtxmath}
\usepackage{color}
\usepackage{dsfont, hyperref, xcolor}
\usepackage{comment}
\usepackage{enumerate}

\newcommand{\abs}[1]{\left\vert#1\right\vert}

\newcommand{\Tr}[1]{\text{Tr}\left\{#1\right\}}

\newcommand{\bra}[1]{\langle#1\vert}
\newcommand{\ket}[1]{\vert#1\rangle}
\newcommand\braket[2]{\langle#1|#2\rangle}

\begin{document}

\title{Quasiprobability distribution of work in the quantum Ising model}

\author{Gianluca~Francica, Luca Dell'Anna}
\address{Dipartimento di Fisica e Astronomia e Sezione INFN, Università di Padova, via Marzolo 8, 35131 Padova, Italy}

\date{\today}

\begin{abstract}
A complete understanding of the statistics of the work done by quenching a parameter of a quantum many-body system is still lacking in the presence of an initial quantum coherence in the energy basis. In this case, the work can be represented by a class of quasiprobability distributions. Here, we try to clarify the genuinely quantum features of the process by studying  the work quasiprobability for an Ising model in a transverse field. We consider both a global and a local quench, by focusing mainly on the thermodynamic limit. We find that, while for a global quench there is a symmetric non-contextual representation with a Gaussian probability distribution of work, for a local quench we can get quantum contextuality as signaled by a negative fourth moment of the work. Furthermore, we examine the critical features related to a quantum phase transition and the role of the initial quantum coherence as useful resource.
\end{abstract}

\maketitle

\section{Introduction}
In the last years out-of-equilibrium processes generated by quenching a parameter of a closed quantum system have been extensively investigated: Outstanding experiments of this kind have been realized with ultra-cold atoms~\cite{Bloch08,Polkovnikov11,Cazalilla11}, and theoretical problems concerning many-body systems have been examined, such as thermalization and integrability~\cite{Polkovnikov11,DAlessio11}, the universality of the dynamics across a critical point~\cite{Dziarmaga10} and the statistics of the work done~\cite{goold18}.
In particular, the work statistics can be described in terms of the two-projective measurement scheme~\cite{Talkner07} if the initial state is incoherent, i.e., there is no initial quantum coherence in the energy basis. In contrast, when the initial state is not incoherent there may not be a
probability distribution for the work done as proven by a no-go theorem~\cite{Perarnau-Llobet17}. This is related to the quantum contextuality as discussed in Ref.~\cite{Lostaglio18}. In simple terms, the problem is similar to looking for a probability distribution in phase space for a quantum particle in a certain quantum state. Since position and momentum are not compatible observables, in general we get a quasiprobability, e.g., the well-known Wigner quasiprobability~\cite{Wigner32}. Concerning the work, which in a thermally isolated quantum system is equal to the energy change of the system, the role of position and momentum is played by the initial and final Hamiltonian of the system. Several attempts have been made to describe the work statistics, among these, quasiprobabilities have been defined in terms of full-counting statistics~\cite{Solinas15} and weak values~\cite{Allahverdyan14}, which can be viewed as particular cases of a more general
quasiprobability introduced in Ref.~\cite{Francica22}. In general, if some fundamental conditions need to be satisfied, the work will be represented by a class of quasiprobability distributions~\cite{Francica222}. Determining the possible representations of the work has a fundamental importance: If there is some quasiprobability that is a non-negative probability, there can be a non-contextual classical representation of the protocol, i.e., the process can be not genuinely quantum.

Here, we focus on the statistics of the work done by quenching a parameter of a many-body system starting from a nonequilibrium state having coherence in the energy basis. Although some investigations on the coherence effects have already been carried out, e.g., in Refs.~\cite{Xu18} and~\cite{Diaz20} the full-counting statistics and weak values quasiprobabilities have been examined, the work statistics still remains rather uninvestigated especially in many-body systems. Thus, after discussing the statistics of work and the quantum contextuality in general in Sec.~\ref{sec. work statistics}, we focus on an Ising model, which we introduce in Sec.~\ref{sec. model}. 
Our aim is to derive some general features of global and local quenches present in the thermodynamic limit thanks to the initial coherence. Furthermore, we are interested to clarify what are the critical features of the work related to a quantum phase transition: Although several studies have been performed for initial incoherent states (e.g., on the large-deviation of work~\cite{gambassi12,Sotiriadis13,Perfetto19} and the Ising model~\cite{Silva08,Dorner12,Fusco14,Russomanno15,Abeling16,Zawadzki23,Fei19,Wang18,Fei20}, just to name a few), also the initial coherence plays a role, as found in Ref.~\cite{Xu18}, which is not entirely clear.
Thus, we focus on a global quench starting from a coherent Gibbs state in Sec.~\ref{sec. global}, where we show that, unlike a system of finite size, in the thermodynamic limit the symmetric quasiprobability representation of the work tends to be non-contextual, in particular we get a Gaussian probability distribution, even if there are also other quasiprobabilities that take negative values. 
In contrast, for a local quench, since the work is not extensive, there are initial states such that all the quasiprobabilities of the class can take negative values as signaled by a negative fourth moment of the work (see Sec.~\ref{sec. local}). Then, these processes remain genuinely quantum also in the thermodynamic limit. Furthermore, we also try to clarify the role of initial quantum coherence as useful resource for the work extraction in Sec.~\ref{sec. cohe}, showing that, even when the protocol tends to be non-contextual, the initial coherence still plays an active role.
In the end, we summarize and discuss further our results in Sec.~\ref{sec. conclusions}.

\section{Work statistics}\label{sec. work statistics}
We consider a quantum quench, so that the system is initially in the state $\rho_0$ and the time evolution is described by the unitary operator $U_{t,0}$ which is generated by the time-dependent Hamiltonian $H(\lambda_t)$ where the control parameter $\lambda_t$ is changed in the time interval $[0,\tau]$. In detail, $U_{t,0}=\mathcal T e^{-i\int_0^t H(\lambda_s) ds}$, where $\mathcal T$ is the time order operator and the Hamiltonian can be expressed as $H(\lambda_t)=\sum_k E_k(\lambda_t) \ket{E_k(\lambda_t)}\bra{E_k(\lambda_t)}$ where $\ket{E_k(\lambda_t)}$ is the eigenstate with eigenvalue $E_k(\lambda_t)$ at the time $t$. For brevity, we define $E_i = E_i(\lambda_0)$ and $E'_k = E_k(\lambda_\tau)$. The average work $\langle w\rangle$ done on the system in the time interval $[0,\tau]$ can be identified with the average energy change
\begin{equation}\label{eq. w ave}
\langle w\rangle = \Tr{(H^{(H)}(\lambda_\tau) -H(\lambda_0) )\rho_0}\,,
\end{equation}
where, given an operator $A(t)$ we define the Heisenberg time evolved operator $A^{(H)}(t) = U_{t,0}^\dagger A(t) U_{t,0}$.
In general, the work performed in the quench can be represented by a quasiprobability distribution of work.
We recall that if some Gleason-like axioms are satisfied (see Ref.~\cite{Francica222} for details), for the events $E\land F$ we get the quasiprobability $v(E,F)=\text{Re}\Tr{E F \rho_0}$, but for more than two events, i.e., for $E\land F \land G \land \cdots$, the quasiprobability is not fixed by the axioms. However, we can associate a quasiprobability of the form $\text{Re}\Tr{E F G \cdots \rho_0}$ to each of all the possible decompositions of the form $E\land F$, $F\land G$, $G \land \cdots$.
By considering this notion of quasiprobability, if we require that (W1) the quasiprobability distribution of work reproduces the two-projective measurement scheme in the case of initial incoherent states (i.e., for states $\rho_0$ such that $\rho_0=\Delta(\rho_0)$, where we have defined the dephasing map $\Delta(\rho_0) = \sum_i \ket{E_i}\bra{E_i} \rho_0\ket{E_i}\bra{E_i}$), (W2) the average calculated with respect to the quasiprobability is equal to Eq.~\eqref{eq. w ave}, and (W3) the second moment is equal to
\begin{equation}\label{eq. w seco}
\langle w^2\rangle = \Tr{(H^{(H)}(\lambda_\tau) -H(\lambda_0) )^2\rho_0}\,,
\end{equation}
the quasiprobability distribution of work belongs to a defined class~\cite{Francica22,Francica222}, i.e., it takes the form
\begin{eqnarray}
\nonumber p_q(w) &=& \sum_{k,j,i} \text{Re}\{\bra{E_i}\rho_0\ket{E_j} \bra{E_j}U^\dagger_{\tau,0}\ket{E'_k}\bra{E'_k}U_{\tau,0}\ket{E_i}\}\\
&& \times  \delta(w-E'_k+qE_i+(1-q)E_j)\,,
\end{eqnarray}
where $q$ is a real parameter. Our aim is to investigate this quasiprobability for a many-body system. We can focus on the characteristic function which is defined as $\chi_q(u) = \langle e^{iuw}\rangle$ and reads
\begin{eqnarray}
\nonumber\chi_q(u) &=& \frac{1}{2}\left( X_q(u)+X_{1-q}(u)\right)\,,
\end{eqnarray}
where we have defined
\begin{equation}\label{eq. X_q}
X_q(u) = \Tr{e^{-iuqH(\lambda_0)}\rho_0 e^{-iu(1-q)H(\lambda_0)} e^{iu H^{(H)}(\lambda_\tau)}}.
\end{equation}
The moments of work are $\langle w^n\rangle = (-i)^n \partial^n_u\chi_q(0)$, and the higher moments for $n>2$ depend on the particular representation. In particular we get
\begin{equation}
\langle w^n \rangle = (-i)^n\partial_u^n\chi_q(0) = \frac{(-i)^n\partial_u^n X_q(0)}{2}+\frac{(-i)^n\partial_u^nX_{1-q}(0)}{2}\,,
\end{equation}
where (see Appendix~\ref{app. w moments})
\begin{eqnarray}
\nonumber(-i)^n\partial_u^n X_q(0) &=& \sum_{k=0}^n (-1)^{n-k}\binom{n}{k}\sum_{l=0}^{n-k}\binom{n-k}{l}q^{n-k-l}(1-q)^l\\
&&\times\Tr{\rho_0 H(\lambda_0)^l(H^{(H)}(\lambda_\tau))^k H(\lambda_0)^{n-k-l}}\,.
\end{eqnarray}
We can consider the problem if there is a classical representation, i.e., if there is a non-contextual hidden variables model which satisfies the conditions about the reproduction of the two-projective-measurement scheme, the average and the second moment.
To introduce the concept of contextuality at an operational level (see, e.g., Refs.~\cite{Lostaglio18,Spekkens08}), we consider a set of preparations procedures $P$ and measurements procedures $M$ with outcomes $k$, so that we will observe $k$ with probability $p(k|P,M)$. We aim to reproduce the statistics by using a set of states $\zeta$ that are random distributed in the set $\mathcal Z$ with probability $p(\zeta|P)$ every time the preparation $P$ is performed. If, for a given $\zeta$, we get the outcome $k$ with the probability $p(k|\zeta,M)$, we are able to reproduce the statistics if
\begin{equation}\label{eq con}
p(k|P,M) = \int_{\mathcal Z} p(\zeta|P) p(k|\zeta,M)d\zeta\,,
\end{equation}
and the protocol is called universally non-contextual if $p(\zeta|P)$ is a function of the quantum state alone, i.e., $p(\zeta|P)=p(\zeta|\rho_0)$, and $p(k|\zeta,M)$ depends only on the positive operator-valued measurement element $M_k$ associated to the corresponding outcome of the measurement $M$, i.e., $p(k|\zeta,M)=p(k|\zeta,M_k)$. In our case, the outcome $k$ corresponds to the work $w_k$, and  if the protocol is non-contextual the work distribution can be expressed as
\begin{equation}\label{eq non cont}
p(w) = \sum_k p(k|P,M) \delta(w-w_k)\,,
\end{equation}
where $p(k|P,M)$ is given by Eq.~\eqref{eq con} with $p(\zeta|P)=p(\zeta|\rho_0)$ and $p(k|\zeta,M)=p(k|\zeta,M_k)$, so that for a negative quasiprobability of work we cannot have a non-contextual protocol.
Thus, a process that cannot be reproduced within any non-contextual protocol will exhibit genuinely non-classical features.
If all the quasiprobabilities in the class take negative values, the protocol is contextual, whereas if there is a quasiprobability which is non-negative, there can be a non-contextual representation.
We recall that for an initial incoherent state $\rho_0=\Delta(\rho_0)$, we get the two-projective measurement scheme that is non-contextual~\cite{Lostaglio18}. In contrast, the presence of initial quantum coherence in the energy basis can lead to a contextual protocol. Let us investigate the effects of the initial quantum coherence by considering a Ising model in a transverse field.

\section{Model}\label{sec. model}
We consider a chain of $L$ spin 1/2 described by the Ising model in a transverse field with Hamiltonian
\begin{equation}
H(\lambda) = - \lambda \sum_{i=1}^L\sigma^z_i - \sum_{i=1}^{L} \sigma^x_i \sigma^x_{i+1}\,,
\end{equation}
where we have imposed periodic boundary conditions $\sigma^\alpha_{L+1}=\sigma^\alpha_1$, and $\sigma^\alpha_i$ with $\alpha=x,y,z$ are the Pauli matrices on the site $i$. We note that the parity $P=\prod_{i=1}^L \sigma^z_i$ is a symmetry of the model, i.e., it commutates with the Hamiltonian. The Hamiltonian can be diagonalized by performing the Jordan-Wigner transformation
\begin{equation}
a_i = \left( \prod_{j<i}\sigma^z_j\right) \sigma^-_i\,,
\end{equation}
where the fermionic operators $a_i$ satisfy the anti-commutation relations $\{a_i,a_j^\dagger\}=\delta_{i,j}$, $\{a_i,a_j\}=0$. We get the Hamiltonian of fermions
\begin{eqnarray}
\nonumber H(\lambda) &=& - \lambda \sum_{i=1}^L(2a^\dagger_ia_i-1) - \sum_{i=1}^{L-1} (a_i^\dagger-a_i)(a_{i+1}+a_{i+1}^\dagger)\\
 &&+ P(a_L^\dagger-a_L)(a_1+a_1^\dagger)\,,
\end{eqnarray}
where the parity reads $P=e^{i\pi N}$ and $N=\sum_{i=1}^L a^\dagger_ia_i$ is the number operator. We consider the projector $P_\pm$ on the sector with parity $P=\pm 1$, then the Hamiltonian reads
\begin{equation}
H(\lambda) =  P_+ H_+(\lambda) P_+ + P_- H_-(\lambda)P_-\,.
\end{equation}
For the sector with odd parity $P=-1$, we get the Kitaev chain
\begin{equation}\label{eq. Kitaev}
H_-(\lambda) = - \lambda \sum_{i=1}^L(2a^\dagger_ia_i-1) - \sum_{i=1}^{L} (a_i^\dagger-a_i)(a_{i+1}+a_{i+1}^\dagger)
\end{equation}
with periodic boundary conditions $a_{L+1}=a_1$. 
We perform a Fourier transform $a_j = 1/\sqrt{L} \sum_k e^{-ikj}a_k$, where $k=2\pi n/L$ with $n=-(L-1)/2, \ldots, (L-1)/2$ for $L$ odd and $n=-L/2+1, \ldots, L/2$ for $L$ even. Thus, the Hamiltonian reads
\begin{equation}
H_-(\lambda) = \sum_k \Psi_k^\dagger \left[-(\lambda+\cos k)\sigma^z + \sin k \sigma^y \right] \Psi_k\,,
\end{equation}
where $\sigma^\alpha$ with $\alpha=x,y,z$ are the Pauli matrices and we have defined the Nambu spinor $\Psi_k = (a_k,a_{-k}^\dagger)^T$. In particular the Hamiltonian can be written as $H_-(\lambda) = \sum_k \Psi_k^\dagger \vec{d}_k\cdot \vec{\sigma} \Psi_k$, which, in the diagonal form, reads
\begin{equation}
H_-(\lambda) =  \sum _k \epsilon_k \left(\alpha_k^\dagger\alpha_k-\frac{1}{2}\right)= \sum _k \epsilon_k \alpha_k^\dagger\alpha_k + E_{-}\,,
\end{equation}
where $E_{-} = -\sum_k \epsilon_k/2$. In detail we have performed a rotation with respect to the $x$-axis with an angle $\theta_k$ between $\vec d_k$ and the $z$-axis, corresponding to the Bogoliubov transformation $\alpha_k = \cos (\theta_k/2) a_k -i \sin(\theta_k/2) a^\dagger_{-k}$, where $\epsilon_k =2 ||\vec d_k||$, or more explicitly,
\begin{equation}
 \epsilon_k =2\sqrt{(\lambda+\cos k)^2+\sin^2 k}\,.
\end{equation}
For the sector with even parity $P=1$, we get the Hamiltonian $H_+(\lambda)$ which is equal to the one in Eq.~\eqref{eq. Kitaev} with antiperiodic boundary conditions $a_{L+1}=-a_1$, thus the only difference is in the momenta $k$ which are $k=2\pi(n-1/2)/L$.
Of course not all eigenstates of the Hamiltonians $H_\pm$ are eigenstates of the Hamiltonian $H$, and their parity needs to be discussed.
Let us consider $L$ even.
Thus, in the even parity sector, $k\in K_+$, for each $k$ there is $-k$, and the eigenstates of the Hamiltonian are the states
\begin{equation}
\alpha^\dagger_{k_1}\cdots \alpha^\dagger_{k_{2m}}\ket{\tilde 0_+}
\end{equation}
where $m$ is an integer, $k_{i}\in K_+$ and $\ket{\tilde 0_+}$ is the vacuum state of $\alpha_k$ with $k\in K_+$. In contrast, in the odd parity sector, $k\in K_-$, for each $k$ there is $-k$ except for $k=0$ and $\pi$. For $\lambda<-1$ we get $\alpha_0=a_0$ and $\alpha_\pi=a_\pi$, for $\lambda>1$ we get $\alpha_0=a_0^\dagger$ and $\alpha_\pi=a_\pi^\dagger$, and for $|\lambda|<1$ we get $\alpha_0=a^\dagger_0$ and $\alpha_\pi = a_\pi$. Then, for $|\lambda|>1$ the vacuum state $\ket{\tilde 0_-}$ of $\alpha_k$ with $k\in K_-$ has even parity, and the eigenstates of the Hamiltonian are the states
\begin{equation}
\alpha^\dagger_{k_1}\cdots \alpha^\dagger_{k_{2m+1}}\ket{\tilde 0_-}
\end{equation}
with $k_{i}\in K_-$. Conversely, for $|\lambda|<1$ the vacuum state $\ket{\tilde 0_-}$ of $\alpha_k$ with $k\in K_-$ has odd parity since has the fermion $a_0$ but not $a_\pi$, and the eigenstates of the Hamiltonian are the states
\begin{equation}
\alpha^\dagger_{k_1}\cdots \alpha^\dagger_{k_{2m}}\ket{\tilde 0_-}
\end{equation}
with $k_{i}\in K_-$. Then, for $|\lambda|<1$ both the states $\ket{\tilde 0_+}$ and $\ket{\tilde 0_-}$ are eigenstates of the Hamiltonian with energies $E_+$ and $E_-$, so that the ground-state is two-fold degenerate in the thermodynamic limit. Thus, at the points $\lambda=\pm 1$ we get a second-order quantum phase transition.

\section{Global quench}\label{sec. global}
\begin{figure}
[t!]
\includegraphics[width=0.79\columnwidth]{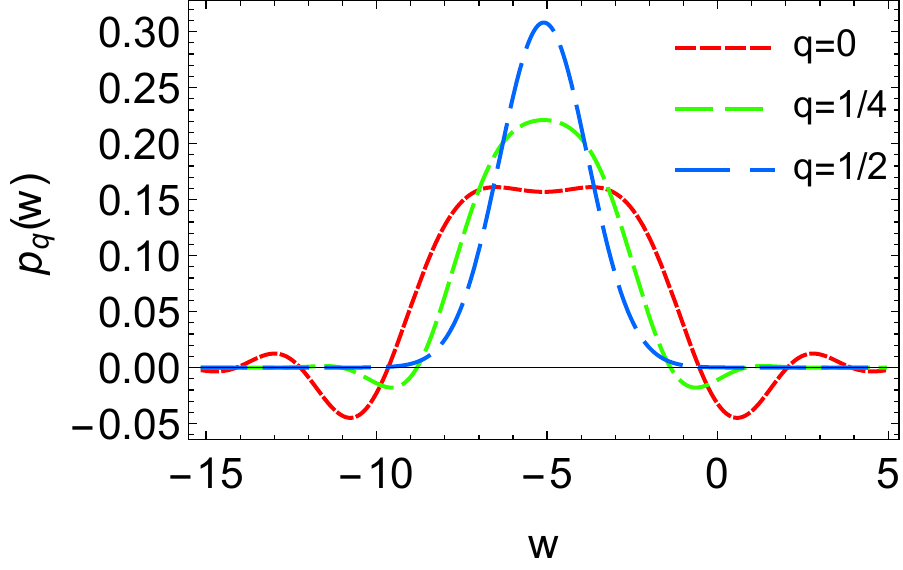}
\caption{ The quasiprobability of work in Eq.~\eqref{eq. gauss complex} for different values of $q$.  We put $L=50$, $\beta=1$, $\lambda_0=0.9$, $\lambda_\tau=1.1$  and $\phi_k=0$.
}
\label{fig:plot pq}
\end{figure}

\begin{figure}
[t!]
\includegraphics[width=0.79\columnwidth]{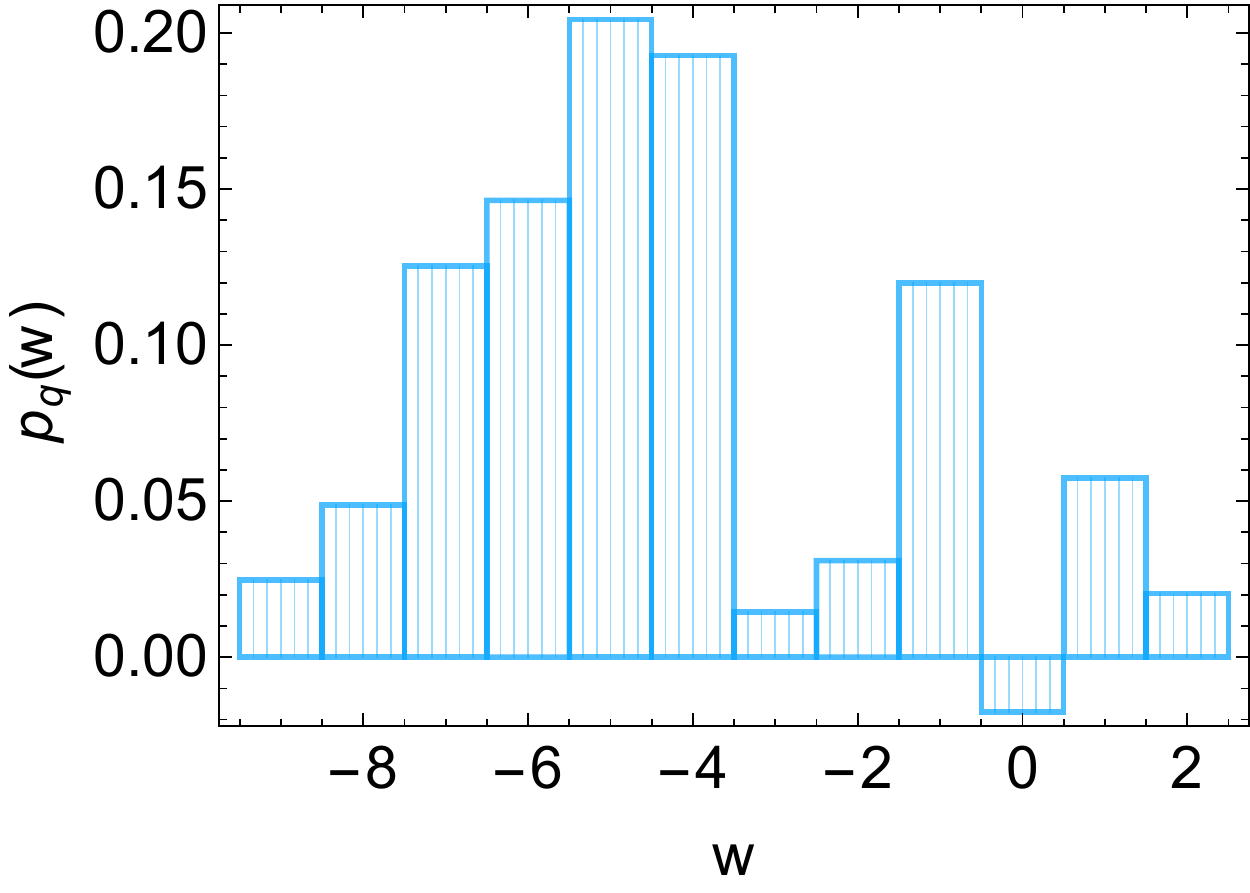}\\
\includegraphics[width=0.79\columnwidth]{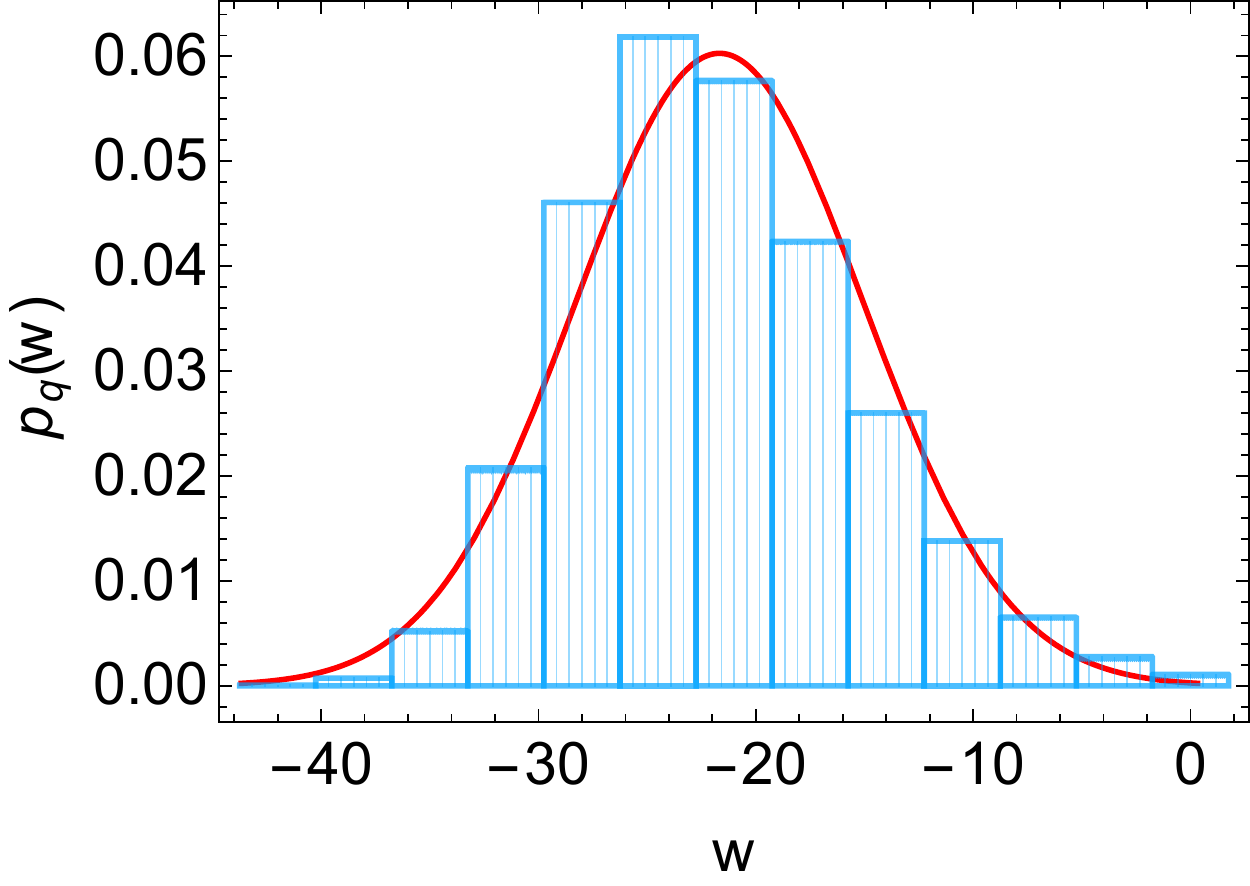}
\caption{ The histogram of the work distribution. We put $L=10$ in the top panel, $L=50$ in the bottom panel, $q=1/2$,  $\beta=1$, $\lambda_\tau=1.5$, $\lambda_0=0.5$ and $\phi_k=\pi/4$. The red line corresponds to the Gaussian distribution probability in Eq.~\eqref{eq. gaussian}. We note that for $L=50$ there is still some skewness. The histograms are calculated by using the characteristic function of Eq.~\eqref{eq. finite size}.
}
\label{fig:plot histo}
\end{figure}

\begin{figure}
[t!]
\includegraphics[width=0.79\columnwidth]{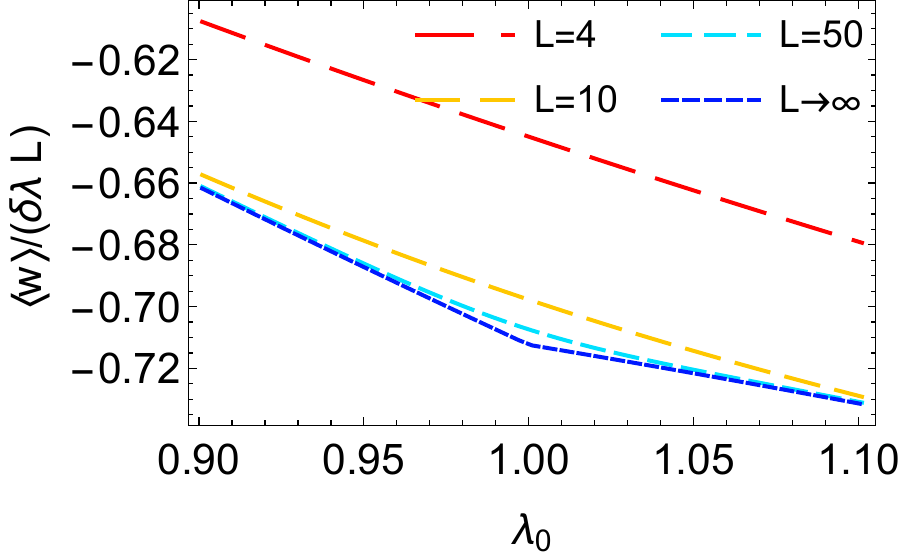}\\
\includegraphics[width=0.79\columnwidth]{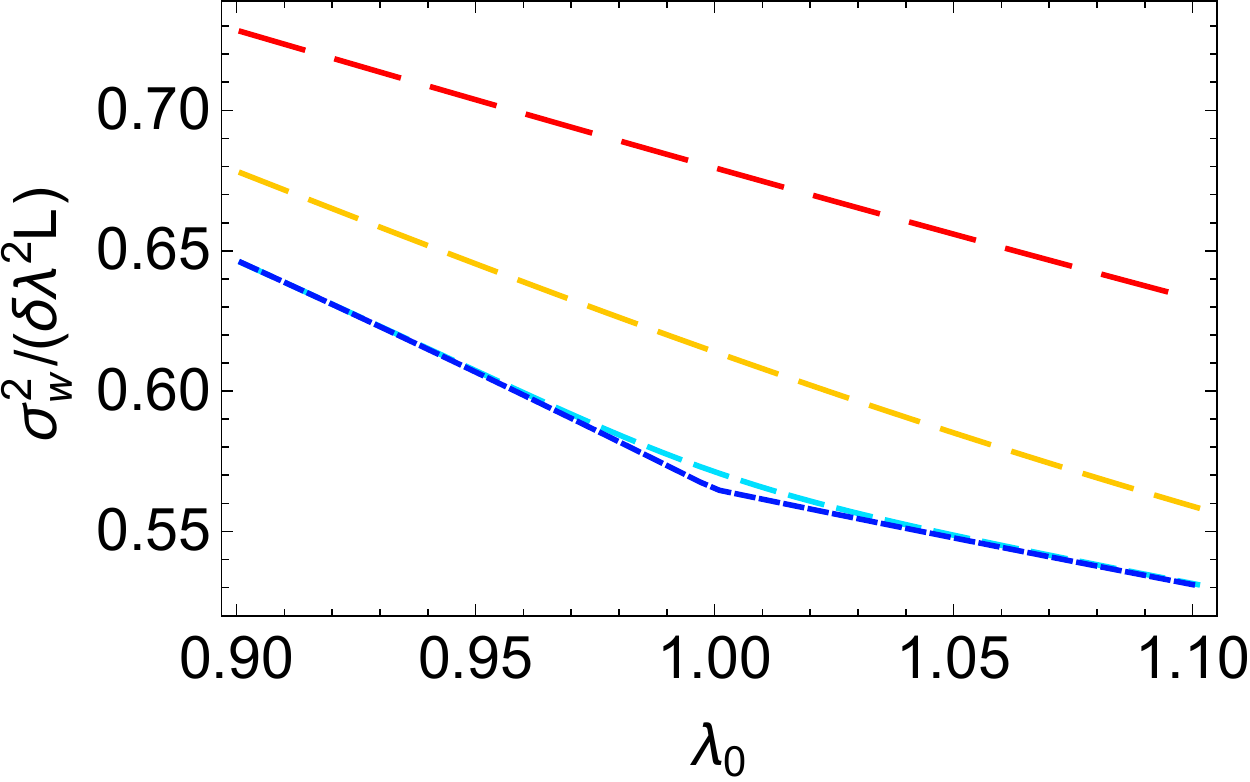}
\caption{ The average work $\bar{w}$ and the variance $\sigma^2_w$ in the function of $\lambda_0$ for different values of $L$. We put $\beta=1$, $\delta \lambda = \lambda_\tau-\lambda_0=0.1$,  and $\phi_k=\pi/4$. The values for finite sizes $L$ are calculated by using the characteristic function of Eq.~\eqref{eq. finite size}.
}
\label{fig:plot vs L}
\end{figure}

We start to focus on a sudden global quench of the transverse field $\lambda$, i.e., $\lambda$ is suddenly changed from the value $\lambda_0$ to $\lambda_\tau$, so that $\tau\to 0$ and $U_{\tau,0}=I$. To investigate the role of initial quantum coherence, we focus on a coherent Gibbs state
\begin{equation}
\ket{\Psi_G(\beta)} =\frac{1}{\sqrt{Z}} \sum_j e^{-\beta E_j/2+i\varphi_j }\ket{E_j}\,,
\end{equation}
where $E_j$ are the eigenenergies of $H(\lambda_0)$, $\varphi_j$ is a phase, $Z=Z(\lambda_0)$ and $Z(\lambda)$ is the partition function defined as $Z(\lambda) = \Tr{e^{-\beta H(\lambda)}}$. Of course, the incoherent part of the state $\ket{\Psi_G(\beta)}$ is $\Delta(\ket{\Psi_G(\beta)}\bra{\Psi_G(\beta)}) = \rho_G(\beta)$, where $\rho_G(\beta)$ is the Gibbs state $\rho_G(\beta)= e^{-\beta H(\lambda_0)}/Z$.
With the aim to calculate the characteristic function for an arbitrary size $L$, from Eq.~\eqref{eq. X_q} by using the relations $\sum_s P_s=I$, $P_s^2=P_s$, $[P_s,H(\lambda)]=0$ and $[P_s,H_\pm(\lambda)]=0$, where $s=\pm$, it is easy to see that
\begin{equation}
 X_q(u) = \sum_s \text{Tr}\big\{e^{-iuqH_s(\lambda_0)}P_s\rho_0 P_s e^{-iu(1-q)H_s(\lambda_0)} e^{iu H_s^{(H)}(\lambda_\tau)}\big\}\,.
\end{equation}
We get $P_s \rho_0 P_s = P_s \rho^s_0$, where for the Gibbs state $\rho^s_0=e^{-\beta H_s(\lambda_0)}/Z$ and for the coherent Gibbs state $\rho^s_0 = \ket{\Psi^s_G}\bra{\Psi^s_G}$. In particular, we get
\begin{equation}\label{eq. psi^s_G}
\ket{\Psi^s_G} = \frac{1}{\sqrt{Z}} \otimes_{k\in K_s} \left( e^{\frac{\beta \epsilon_k}{4}}\ket{\tilde 0_k}+e^{-\frac{\beta \epsilon_k}{4}+i\phi_k}\ket{\tilde 1_k}\right)\,,
\end{equation}
where we consider a phase such that $\phi_{-k}=\phi_k$, with $\ket{\tilde n_k} = (\alpha^\dagger_k)^{n_k}\ket{\tilde 0_k}$, where $\epsilon_k=\epsilon_k(\lambda_0)$, $\alpha_k=\alpha_k(\lambda_0)$ and $\ket{\tilde 0_k}$ is the vacuum state for the fermion $\alpha_k$. As shown in Appendix~\ref{app. chara}, we get
\begin{equation}\label{eq. finite size}
X_q(u) = \frac{1}{2}\sum_s X^s_q(u) + \eta_s X'^s_q(u)\,,
\end{equation}
where we have defined $\eta_s = s \bra{\tilde 0_s}e^{i\pi N} \ket{\tilde 0_s}$ which is $\eta_+ =1$ and $\eta_-=-1$ for $|\lambda_0|>1$ and $\eta_-=1$ for $|\lambda_0|<1$, and
\begin{equation}\label{eq. eqeq}
X^s_q(u) = \frac{1}{Z}\prod_{k\in K_s : k\geq 0} X^{(k)}_q(u)\,.
\end{equation}
In detail, for $k> 0$ and $k\neq \pi$, we get
\begin{equation}
X^{(k)}_q(u) = X^{(k),th}_q(u) +X^{(k),coh}_q(u)\,,
\end{equation}
where $X^{(k),th}_q(u)$ is the incoherent contribution, which reads
\begin{eqnarray}
\nonumber X^{(k),th}_q(u) &=& 2 \bigg( \cos((u-i\beta)\epsilon_k) \cos(u\epsilon'_k) + \sin((u-i\beta)\epsilon_k)\\
&& \times\sin(u\epsilon'_k)\hat{d}_k\cdot \hat{d}'_k +1\bigg)
\end{eqnarray}
and $X^{(k),coh}_q(u)$ is the coherent contribution, which reads
\begin{equation}
X^{(k),coh}_q(u) = - 2i\sin(u\epsilon'_k)\sin(u(2q-1)\epsilon_k-2\phi_k)(\hat{d}_k\times \hat{d}'_k)_x\,,
\end{equation}
where, for brevity we have defined $\epsilon'_k=\epsilon_k(\lambda_\tau)$, $\vec d_k=\vec d_k(\lambda_0)$ and $\vec d'_k=\vec d_k(\lambda_\tau)$. Furthermore, we have
\begin{equation}
X'^s_q(u) = \frac{1}{Z}\prod_{k\in K_s : k\geq 0} X'^{(k)}_q(u)
\end{equation}
with
\begin{equation}
X'^{(k)}_q(u)=  X^{(k)}_q(u)-4\,.
\end{equation}
In contrast, for $k=0$ and $k=\pi$, we get
\begin{eqnarray}
X^{(0,\pi)}_q(u) &=& 2\cosh\left(\frac{\beta \epsilon_{0,\pi} - i u (s_{0,\pi}\epsilon'_{0,\pi}-\epsilon_{0,\pi})}{2}\right)\,,\\
X'^{(0,\pi)}_q(u) &=& 2\sinh\left(\frac{\beta \epsilon_{0,\pi} - i u (s_{0,\pi}\epsilon'_{0,\pi}-\epsilon_{0,\pi})}{2}\right)\,,
\end{eqnarray}
where $s_{\pi}=-1$ if $|\lambda_0|<1$ and $\lambda_\tau>1$ or $|\lambda_\tau|<1$ and $\lambda_0>1$, otherwise $s_{\pi}=1$, and $s_{0}=-1$ if $|\lambda_0|<1$ and $\lambda_\tau<-1$ or $|\lambda_\tau|<1$ and $\lambda_0<-1$, otherwise $s_{0}=1$, while the partition function is
\begin{equation}
Z = \frac{1}{2}\sum_s \prod_{k\in K_s} 2\cosh(\beta \epsilon_k/2) + \eta_s \prod_{k\in K_s} 2\sinh(\beta \epsilon_k/2)\,.
\end{equation}
If the initial quantum coherence does not contribute, i.e., $X^{(k),coh}_q(u)=0$, we get $X^{(k)}_q(u) = X^{(k),th}_q(u)$ and the characteristic function is the one of the initial Gibbs state $\rho_G(\beta)$.
We get $X^{(k),coh}_q(u)=0$ for $q=1/2$  and $\phi_k= n \pi/2$, and in this case the quasiprobability is non-negative, in particular it is equivalent to the two-projective-measurement scheme which is non-contextual.
For $q=1/2$  the initial quantum coherence contributes only for $\phi_k\neq n \pi/2$ with $n$ integer. In this case the quasiprobability can take negative values. However, in the thermodynamic limit the negativity of the quasiprobability is always subdominant for $q=1/2$, and we get a Gaussian probability distribution of work. 
To prove it, we note that in the thermodynamic limit we get  $Z= \prod_{k\in K_+} Z_k$ with $Z_k=2\cosh(\beta \epsilon_k/2)$, then
\begin{equation}\label{eq. XQQ}
X_q(u) = \prod_{k\in K_+ : k>0} \frac{ X^{(k)}_q(u)}{Z_k^2}\,.
\end{equation}
Basically, in the thermodynamic limit the model is equivalent to the system of fermions with Hamiltonian $H_+$.
 Thus, we can write
\begin{equation}\label{eq. Xq L}
X_q(u)  = e^{L g_q(u)}\,,
\end{equation}
where $g_q(u)$ is intensive, so that the work is extensive, i.e., $\langle w^n \rangle \sim L^n$. In particular, for the initial coherent Gibbs state under consideration, $g_q(u)$ explicitly reads
\begin{equation}
g_q(u) = \frac{1}{2\pi}\int_0^\pi \ln\left(\frac{X^{(k)}_q(u)}{Z_k^2}\right) dk\,.
\end{equation}
Then, if Eq.~\eqref{eq. Xq L} holds, regardless of the explicit form of the intensive function $g_q(u)$, as $L\to\infty$ we can consider
\begin{equation}\label{eq. X_q limit L}
X_q(u) \sim e^{L \left(\partial_u g_q(0) u + \frac{1}{2}\partial_u^2 g_q(0)u^2\right)}
\end{equation}
since in the calculation of the Fourier transform of $X_q(u)$ the dominant contribution of the integral is near $u=0$, so that we can expand $g_q(u)$ in Taylor series about $u=0$, and thus the neglected terms in Eq.~\eqref{eq. X_q limit L} do not contribute in the asymptotic formula of the quasiprobability $p_q(w)$.
We note that, although the characteristic function $\chi_q(u)$ depends on $q$, the first two moments do not depend on $q$. In particular, we note that the relative fluctuations of work scale as $\sigma_w/\langle w\rangle \sim 1/\sqrt{L}$, where we have defined the variance $\sigma_w^2=\langle w^2\rangle - \langle w\rangle^2$.
By noting that $\partial_u g_{q}(0)$ does not depend on $q$ and $\partial_u^2 g_{1-q}(0)=\partial_u^2 g^*_q(0)$, we get the quasiprobability of work
\begin{equation}\label{eq. gauss complex}
p_q(w) \sim \frac{1}{\sqrt{2\pi}}\text{Re}\left( \frac{e^{-\frac{(w -\bar{w})^2}{2 v_q}}}{\sqrt{v_q}}\right)\,,
\end{equation}
where $\bar{w}=-i\partial_u g_{q}(0)L$ and $v_q=-\partial_u^2 g_{q}(0)L$. In particular the average work  is $\langle w \rangle = \bar{w}$  and the variance $\sigma_w^2$ is the real part of $v_q$, i.e., $v_q = \sigma_w^2 +i r_q$. As shown in Fig.~\ref{fig:plot pq}, for $q\neq 1/2$ the asymptotic formula of the quasiprobability can take negative values due to the presence of the imaginary part $r_q$.
In contrast, for $q=1/2$, we get $\chi_{1/2}(u) = X_{1/2}(u)$, from which $\sigma_w^2=-\partial_u^2 g_{1/2}(0)L$, i.e., $r_{1/2}=0$ and thus we get the Gaussian probability distribution
\begin{equation}\label{eq. gaussian}
p_{1/2}(w)\sim  \frac{e^{-\frac{(w -\bar{w})^2}{2 \sigma^2_w}}}{\sqrt{2\pi}\sigma_w}\,.
\end{equation}
It is worth noting that the protocol tends to be non-contextual. To prove it, we consider the operator $\Delta H = H^{(H)}(\lambda_\tau) -H(\lambda_0)$, and the probability distribution
\begin{equation}
p(\Delta E ) = \sum_{\mu} \bra{\Delta E_\mu} \rho_0 \ket{\Delta E_\mu}\delta(\Delta E-\Delta E_\mu)\,,
\end{equation}
where $\ket{\Delta E_\mu}$ is the eigenstate of $\Delta H$ with eigenvalue $\Delta E_\mu$. Of course $p(\Delta E )$ is non-contextual, and it is easy to see that $p_{1/2}(w)\sim p(w) $ as $L\to \infty$.
Thus, the work tends to be an observable with respect to the non-contextual symmetric representation, differently from finite sizes, where it is not, as was originally noted for incoherent initial states in Ref.~\cite{Talkner07}.
In particular for the quench considered, we have $\Delta H = (\lambda_\tau-\lambda_0)S_z$, where $S_z=\sum_{j=1}^L \sigma^z_j$, so that the symmetric representation for $q=1/2$ tends to be equivalent to the distribution probability of the transverse magnetization $S_z$.
We emphasize that for small sizes $L$ the quasiprobability at $q=1/2$ can take negative values, but for large $L$ it is well described by the Gaussian probability distribution in Eq.~\eqref{eq. gaussian} (see Fig.~\ref{fig:plot histo}).
 We note that for an arbitrary initial state the representation for $q=1/2$ is still non-contextual (see Appendix~\ref{app. super}).
In general, the negativity of the quasiprobability $p_q(w)$ can be characterized by the integral
\begin{equation}
\mathcal N \equiv \int |p_q(w)|dw\,,
\end{equation}
which is equal to one if $p_q(w)\geq 0$. In our case, $\mathcal N \sim \frac{(\sigma_w^4+r^2_q)^{\frac{1}{4}}}{\sigma_w}$, so that $\mathcal N=1$ implies that $r_q=0$ and thus $p_q(w)\geq 0$.
We note that $\mathcal N=1$ implies in general that $p_q(w)\geq 0$ (see Appendix~\ref{app. negativity}).
In the end we note that the effects related to the negativity of the quasiprobability start to affect the statistics from the fourth moment, which reads $\langle w^4 \rangle  \sim\bar{w}^4 + 6 \bar{w}^2 \sigma^2_w + 3 \sigma^4_w- 3 r_q^2$. In contrast the first three moments do not depend on $r_q$, explicitly they read $\langle w\rangle = \bar{w}$, $\langle w^2\rangle = \bar{w}^2+\sigma_w^2$ and $\langle w^3\rangle \sim \bar{w}^3+3\bar{w}\sigma_w^2$. In particular, the kurtosis is $\text{Kurt} \equiv \langle (w-\langle w\rangle)^4\rangle/\sigma^4_w \sim 3 - 3 r^2_q/\sigma_w^4$ which is always smaller than 3 if $r_q\neq0$, i.e., the distribution is more `flat' than the normal one. We note that if $\bar{w}\neq 0$, since $\bar{w}\sim L$ and $\sigma^2_w \sim L$, the fourth moment is always positive. On the other hand, for $\bar{w}= 0$, the fourth moment reads $\langle w^4 \rangle  \sim 3 \sigma^4_w- 3 r_q^2$ and becomes negative for $r_q>\sigma^2_w$, so that in this regime the negativity for $q\neq 1/2$ will be strong.
To conclude our investigation concerning the global quench, we note that the average work reads
\begin{equation}\label{eq. wbar}
\bar{w} =\frac{(\lambda_0-\lambda_\tau)L}{\pi}\int_0^\pi \frac{(\lambda_0+\cos k)\sinh(\beta \epsilon_k) + \sin k \sin(2\phi_k)}{ \epsilon_k\cosh^2(\beta \epsilon_k/2)}dk
\end{equation}
and the variance reads
\begin{eqnarray}\label{eq. sigmaw}
\nonumber \sigma_w^2 &=& \frac{(\lambda_0-\lambda_\tau)^2L}{\pi}\int_0^\pi \frac{1}{\cosh^4(\beta \epsilon_k/2)}\bigg( \cosh^2(\beta \epsilon_k/2) \\
\nonumber &&\times\cosh(\beta\epsilon_k) - \frac{2}{\epsilon_k^2} \big( \sin k \sin(2\phi_k) + (\lambda_0+\cos k)\\
&&\times \sinh(\beta \epsilon_k)\big)^2 \bigg)dk\,.
\end{eqnarray}
Both $\bar{w}$ and $\sigma_w^2$ are not regular at $|\lambda_0|=1$ for $\phi_k=\phi\neq n\pi/2$ due to the presence of a quantum phase transition (see Fig.~\ref{fig:plot vs L}).
Furthermore, concerning the negativity of the quasiprobability of work, we have
\begin{equation}
r_q = \frac{2(1-2q)(\lambda_\tau-\lambda_0)L}{\pi} \int_0^\pi \frac{\sin k \cos(2\phi_k)}{\cosh^2(\beta\epsilon_k/2)}dk\,,
\end{equation}
which is regular. We deduce that the protocol admits a non-contextual description, i.e., $r_q=0$, for any $q$ and $\phi_k = (2n+1)\pi/4$ or for $q=1/2$.
In the end, to investigate the critical features of the work which can be related to the presence of the quantum phase transition, we introduce the energy scale $J$ such the Hamiltonian reads
\begin{equation}
H_J(\lambda) = - J \lambda \sum_{i=1}^L\sigma^z_i - J\sum_{i=1}^{L} \sigma^x_i \sigma^x_{i+1}\,.
\end{equation}
We focus on $\lambda_0\approx 1$ and we start to consider the average work given by Eq.~\eqref{eq. wbar} multiplied by $J$. Then, we change variable $k'=\pi-k$ in the integral and we define $\kappa = k'/a$, and the renormalized couplings $J=c/(2a)$ and $\lambda_0=1-mca$. In the scaling limit $a\to 0$, we get
\begin{equation}\label{eq. wbar cont}
\bar{w} \sim \frac{J(\lambda_0-\lambda_\tau)aL}{2\pi}\int_0^{\frac{\pi}{a}} \frac{\kappa \sin(2\phi_\pi)-cm\sinh(\beta c \omega_\kappa)}{\omega_\kappa\cosh^2(\beta c \omega_\kappa/2)}d\kappa\,,
\end{equation}
where $\omega_\kappa = \sqrt{\kappa^2+c^2m^2}$. We note that the integral extended to the interval $[0,\infty)$ does not converge. Thus the integral is not determined only by small $\kappa$, and the behavior is not universal. Similarly, concerning the variance $\sigma_w^2$, the integral extended to the interval $[0,\infty)$ does not converge, so that it is not universal. The coherent contribution to the average work is defined as
\begin{equation}
\bar{w}_{coh} = \bar{w} - \bar{w}_{th}\,,
\end{equation}
where $\bar{w}_{th}$ is the average work corresponding to the initial state $\rho_0=\rho_G(\beta)$. Then, the coherent contribution is given by the term proportional to $ \sin(2\phi_\pi)$ in Eq.~\eqref{eq. wbar cont}, i.e.,
\begin{equation}\label{eq. wbar cont cohe}
\bar{w}_{coh} \sim \frac{J(\lambda_0-\lambda_\tau)aL}{2\pi}\int_0^{\frac{\pi}{a}} \frac{\kappa \sin(2\phi_\pi)}{\omega_\kappa\cosh^2(\beta c \omega_\kappa/2)}d\kappa\,.
\end{equation}
In this case we can  extend the integral to the interval $[0,\infty)$, so that the coherent contribution $\bar{w}_{coh}$ is described by the continuum  model, in this sense it is a universal feature. From Eq.~\eqref{eq. wbar cont cohe}, by noting that
\begin{equation}
\int_0^\infty \frac{y}{\sqrt{1+y^2}\cosh^2(x\sqrt{1+y^2}/2)}dy= \frac{4}{(1+e^{|x|})|x|}\,,
\end{equation}
the coherent contribution to the average work can be expressed as
\begin{equation}
\bar{w}_{coh} \sim \frac{(\lambda_0-\lambda_\tau)\sin(2\phi_\pi)L}{\pi\beta} g_{FD}(\beta m c^2)\,,
\end{equation}
where we have defined the Fermi-Dirac distribution $g_{FD}(x)=1/(1+e^{|x|})$ and $ m c^2=2J(1-\lambda_0)$.
In the end, let us consider the limit of high temperatures $\beta\to 0$, so that  we get
\begin{eqnarray}\label{eq. g_q}
\nonumber  g_q(u)&=&  \frac{1}{2\pi}\int_0^\pi \ln\frac{1}{2}\bigg( \cos(u\epsilon_k) \cos(u\epsilon'_k) + \sin(u\epsilon_k)\sin(u\epsilon'_k)\\
\nonumber && \times\hat{d}_k\cdot \hat{d}'_k  +1 - i\sin(u\epsilon'_k)\sin(u(2q-1)\epsilon_k-2\phi_k)\\
&& \times (\hat{d}_k\times \hat{d}'_k)_x\bigg)dk\,.
\end{eqnarray}
For $\phi_k=\phi$, we get the closed form of the derivatives
\begin{eqnarray}
\partial_u g_q(0)&=& -\frac{i(\lambda_\tau-\lambda_0)}{2\pi |\lambda_0|} \sin(2\phi)(1+|\lambda_0| - |1-|\lambda_0||)\,,\\
\nonumber \partial_u^2 g_q(0) &=& -(\lambda_\tau-\lambda_0)^2\bigg( 1 - \frac{1}{8\lambda_0^2}\left(1+\lambda_0^2-(1+|\lambda_0|)|1-|\lambda_0||\right)\\
&&\times\sin^2(2\phi) \bigg)-\frac{4i}{\pi}(\lambda_\tau-\lambda_0)(1-2q)\cos(2\phi)\,,
\end{eqnarray}
from which it is evident that the work statistics is not regular at $|\lambda_0|=1$  for $\phi\neq n\pi/2$. Of course in this limit we can extract the work $W_{ex} = -\langle w \rangle $, equals to
\begin{equation}\label{eq. work extra zero}
W_{ex}= \frac{(\lambda_\tau-\lambda_0)L}{2\pi |\lambda_0|} \sin(2\phi)(1+|\lambda_0| - |1-|\lambda_0||)\,,
\end{equation}
only because of the presence of the initial coherence, otherwise for an initial Gibbs state we will get $\langle w \rangle =0$. 


\section{Local quench}\label{sec. local}
\begin{figure}
[t!]
\includegraphics[width=0.79\columnwidth]{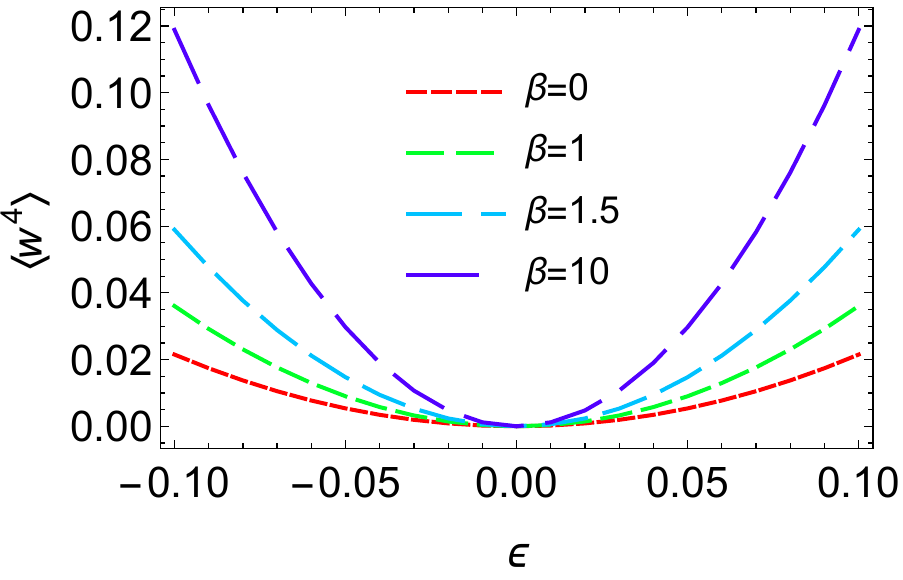}\\
\includegraphics[width=0.79\columnwidth]{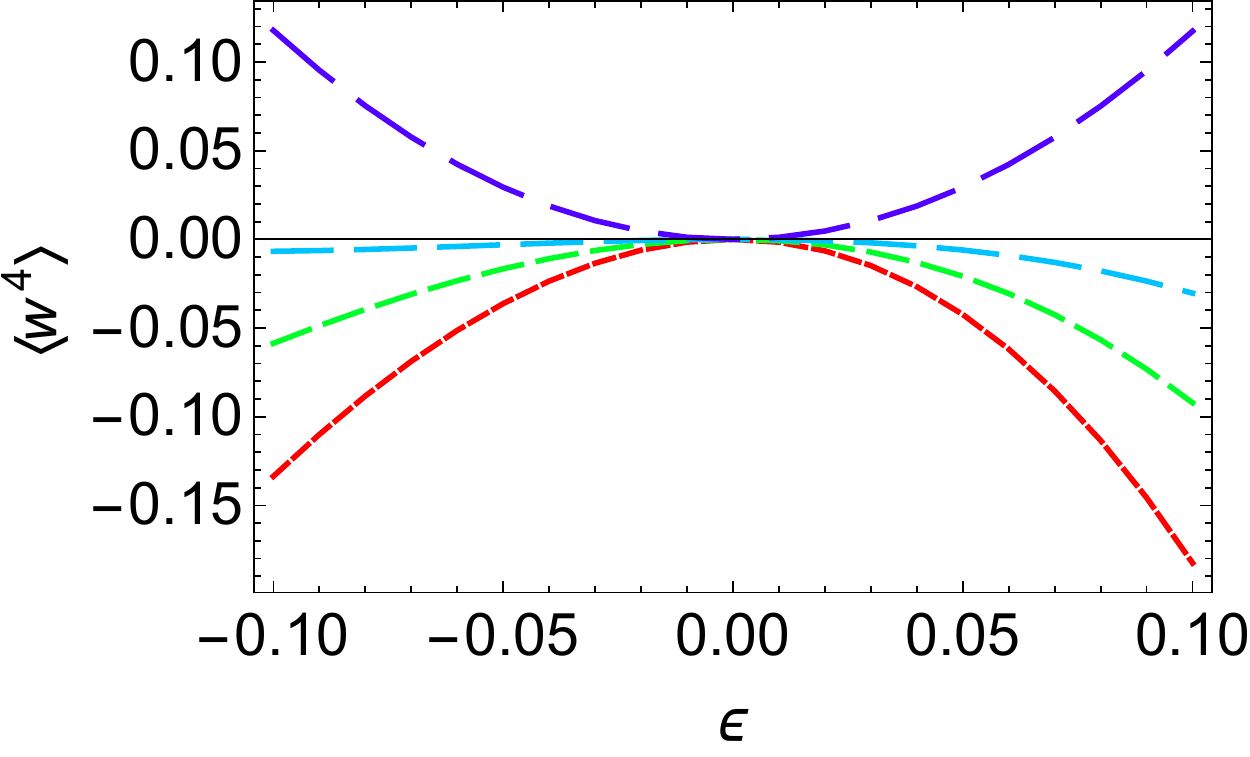}
\caption{ The fourth moment of work $\langle w^4\rangle$ in the function of the local field $\epsilon$ for $q=1/2$ for the states $\ket{\Psi_1(\beta)}$ (top panel) and $\ket{\Psi_2(\beta)}$ (bottom panel). The curves for other values of $q\in[0,1/2]$ are not distinguishable by eye from the one for $q=1/2$. We put $L=50$, $\lambda_0=1$, $\phi_k=\phi_{-k}=\pi$ and $\phi_k=\phi_{-k}=0$ for $n$ odd and even, where $k=2\pi(n-1/2)/L$.
}
\label{fig:plot 4th}
\end{figure}

Things change drastically when the work is non-extensive, e.g., for a local quench. We focus on the case of a sudden quench in the transverse field, i.e., the initial Hamiltonian is $H=H(\lambda_0)$ and we perform a sudden quench of the transverse field in a site $l$, so that the final Hamiltonian is $H'=H(\lambda_0)-\epsilon\sigma^z_l$. Since we are interested only to large sizes $L$, we describe the model with the corresponding fermionic Hamiltonian $H_+$.
Here we are interested to investigate how contextuality can emerge in a local quench, thus we focus on the states $\ket{\Psi_1(\beta)}$ and $\ket{\Psi_2(\beta)}$, which are defined as
\begin{equation}
\ket{\Psi_1(\beta)} = \frac{e^{\frac{\beta}{4}\sum_k\epsilon_k}}{\sqrt{Z_1}}\exp\left(\sum_ke^{-\frac{\beta \epsilon_k}{2}+i\phi_k}\alpha_k^\dagger\right)\ket{\tilde 0_+}
\end{equation}
and
\begin{eqnarray}
\nonumber \ket{\Psi_2(\beta)} &=& \frac{e^{\frac{\beta}{4}\sum_k\epsilon_k}}{\sqrt{Z_2}}\bigg(1+\sum_ke^{-\frac{\beta \epsilon_k}{2}+i\phi_k}\alpha_k^\dagger \\
&& + \frac{1}{2} \sum_{k,k'}s_{k,k'}e^{-\frac{\beta (\epsilon_k+\epsilon_{k'})}{2}+i(\phi_k+\phi_{k'})}\alpha_k^\dagger \alpha_{k'}^\dagger \bigg)\ket{\tilde 0_+}\,,
\end{eqnarray}
where $s_{k,k'}=1$ if $k>k'$,  $s_{k,k'}=-1$ if $k<k'$ and  $s_{k,k}=0$, and $Z_1$ and $Z_2$ are normalization factors such that $Z\sim Z_2\sim Z_1$ as $\beta \to \infty$. Indeed, $\ket{\Psi_G(\beta)}\sim \ket{\Psi_2(\beta)} \sim \ket{\Psi_1(\beta)}$ as $\beta \to \infty$. In general, for these initial states, the function $X_q(u)$ can be calculated with the help of Grassmann variables (see Appendix~\ref{app. quadratic form}).
While for the initial state $\ket{\Psi_1(\beta)}$, we find that the fourth moment of work is positive, for the initial state $\ket{\Psi_2(\beta)}$, we find that the fourth moment of work can be negative for $\beta$ small enough (see Fig.~\ref{fig:plot 4th}).
This suggests that to get a contextual protocol with a negative fourth moment we need to start from an initial state which involves at least couples of quasiparticles, e.g., such as $\ket{\Psi_2(\beta)}$. This result is corroborated by considering states like $\ket{\Psi_1(\beta)}$ but with random coefficients instead of $e^{-\frac{\beta \epsilon_k}{2}+i\phi_k}$, for which we get a non-negative fourth moment for the local quench.

\section{Initial quantum coherence}\label{sec. cohe}
To conclude we investigate further the role of initial coherence by focusing on an initial state $\rho_0$ with a thermal incoherent part, i.e., $\Delta(\rho_0)=\rho_G(\beta)$. In general, we have the equality (see Ref.~\cite{Francica22})
\begin{equation}\label{eq. fluc theo}
\langle e^{-\beta w-C} \rangle =e^{-\beta \Delta F}\,,
\end{equation}
where $\Delta F=F(\lambda_\tau)-F(\lambda_0)$ is the change in the equilibrium free energy, where $F(\lambda)=-\beta^{-1} \ln Z(\lambda)$, and $C$ is the random quantum coherence that has the probability distribution
\begin{equation}
p_c(C) = \sum_{i,n} R_n \abs{\braket{E_i}{R_n}}^2\delta(C+\ln\bra{E_i}\rho_0\ket{E_i}-\ln R_n)\,,
\end{equation}
where we have considered the decomposition $\rho_0 = \sum_n R_n \ket{R_n}\bra{R_n}$. In detail, the average of $C$ is the relative entropy of coherence $\langle C\rangle = S(\Delta(\rho_0))-S(\rho_0)$, where $S(\rho)$ is the von Neumann entropy defined as $S(\rho)=-\Tr{\rho\ln\rho}$, and we have the equality $\langle e^{-C}\rangle =1$. In particular, from Eq.~\eqref{eq. fluc theo}, we get the inequality $\langle w \rangle  \geq \Delta F - \beta^{-1}\langle C \rangle $, and we note that Eq.~\eqref{eq. fluc theo} reduces to the Jarzynski equality~\cite{Jarzynski97} $\langle e^{-\beta w} \rangle =e^{-\beta \Delta F}$ when $\rho_0=\rho_G(\beta)$.
From Eq.~\eqref{eq. fluc theo} we get
\begin{equation}\label{eq. cumu series}
\Delta F = \beta^{-1} \sum_{n=1}^{\infty} \frac{(-1)^{n+1}}{n!}\kappa_n(s)\,,
\end{equation}
where $\kappa_n(s)$ is the nth cumulant of $s=\beta w+C$ which of course it can be expressed in terms of expectation values of work and coherence: The cumulants $\kappa_n(C)$ of $C$ cancel in the sum due to the equality $\langle e^{-C}\rangle=1$, and only  work cumulants $\kappa_n(w)$ (e.g., the variance $\sigma_w^2$) and correlation terms (e.g., the covariance $\sigma_{w,C}=\langle w C \rangle - \langle w\rangle \langle C\rangle$) are present. For instance, if work and coherence are uncorrelated, we get $\kappa_n(s)=\beta^n \kappa_n(w)+\kappa_n(C)$ and so $\Delta F =  \sum_{n=1}^{\infty} (-1)^{n+1}\beta^{n-1}\kappa_n(w)/n! $ and the coherence does not appear. If we consider a Gaussian probability distribution for the random variable $s$, we get
\begin{equation}\label{eq. work extra}
\Delta F  = \langle w \rangle - \frac{\beta \sigma^2_w}{2} - \sigma_{w,C}\,.
\end{equation}
For a given free energy change $\Delta F$, from Eq.~\eqref{eq. work extra} we see that the average work extracted $W_{ex}=-\langle w\rangle $ in the process increases as the fluctuation of work becomes weak, i.e., the variance $\sigma^2_w$ decreases, and the work and coherence become strongly negative correlated, i.e., $\sigma_{w,C}<0$, which clarifies the role of initial quantum coherence as useful resource.
However, we note that Eq.~\eqref{eq. work extra} cannot be exactly satisfied for a global quench because we have to take into account also higher work cumulants and correlations which will contribute to the series in Eq.~\eqref{eq. cumu series} due to large deviations. 
In particular, if we focus on the high temperature limit $\beta\to 0$,  Eq.~\eqref{eq. cumu series} reduces to
\begin{equation}\label{eq. w cumu}
\langle w \rangle = \Delta F + \sum_{k=1}^\infty \frac{i^{k+1}}{k!}\partial^{k}_t\partial_u G(0,0)\,,
\end{equation}
where we have defined the function $G(u,t)=\ln  \langle e^{iuw+itC}\rangle$. The derivatives are correlation terms, e.g., $\partial_t\partial_u G(0,0) = - \sigma_{w,C}$, $\partial^2_t\partial_u G(0,0)= 2i\langle C \rangle\sigma_{w,C}- i \sigma_{w,C^2}$ and $\partial^3_t \partial_u G(0,0) = 3(2 \langle C\rangle^2 - \langle C^2\rangle ) \sigma_{w,C}-3 \langle C \rangle \sigma_{w,C^2}+\sigma_{w,C^3}$.
For the initial state $\rho_0 = \eta \ket{\Psi_G(0)}\bra{\Psi_G(0)} + (1-\eta)\rho_G(0)$, we get the characteristic function of the coherence (see Appendix~\ref{app. ini cohe})
\begin{equation}\label{eq. ini cohe}
\langle e^{i t C} \rangle = D^{it}\left( \left(\eta+\frac{1-\eta}{D}\right)^{it+1}+(D-1)\left(\frac{1-\eta}{D}\right)^{it+1}\right)\,,
\end{equation}
where $D$ is the dimension of the Hilbert space.
Furthermore, by considering
\begin{eqnarray}\label{eq. joint Cw}
\nonumber \langle e^{iuw+itC}\rangle &=& \text{Tr}\{\rho_0 e^{i t \ln \rho_0} e^{-iuH/2-i t \ln\Delta(\rho_0)/2}e^{iu H'} \\
&&\times   e^{-iuH/2-i t \ln\Delta(\rho_0)/2}\}\,,
\end{eqnarray}
where for brevity we have defined $H=H(\lambda_0)$ and $H'=H^{(H)}(\lambda_\tau)$, we get
\begin{equation}\label{eq. deri}
-i\partial_u G(0,t)= \frac{\left(\eta+\frac{1-\eta}{D}\right)^{it+1}w_1+(D-1)\left(\frac{1-\eta}{D}\right)^{it+1}w_2}{\left(\eta+\frac{1-\eta}{D}\right)^{it+1}+(D-1)\left(\frac{1-\eta}{D}\right)^{it+1}}\,,
\end{equation}
where $w_1=\bra{\Psi_G(0)}(H'-H)\ket{\Psi_G(0)}$ is the average work done starting from the coherent Gibbs state, which can be expressed as $w_1 = \left( \langle w \rangle - (1-\eta)\Delta F\right)/\eta$, and $w_2 = D \Delta F - w_1$, where $\Delta F = \Tr{H(\lambda_\tau)-H(\lambda_0)}/D$. Thus, the terms in Eq.~\eqref{eq. w cumu} can be obtained by calculating the derivatives of Eq.~\eqref{eq. deri} with respect to $t$.
We note that for the Ising model we get $\Delta F = 0$, so that in this limit the work extracted, i.e., Eq.~\eqref{eq. work extra zero} multiplied by $\eta$, completely comes from the correlations between work and coherence. Of course, the same situation occurs for a cyclic change of any Hamiltonian, i.e., such that $H(\lambda_\tau)=H(\lambda_0)$.

\section{Conclusions}\label{sec. conclusions}
We investigated the effects of the initial quantum coherence in the energy basis to the work done by quenching a transverse field of a one-dimensional Ising model. The work can be represented by considering a class of quasiprobability distributions. To study how the work statistics changes with the increasing of the system size, we calculated the exact formula of the characteristic function of work for an arbitrary size by imposing periodic boundary conditions. Then, we focused on the thermodynamic limit, and we showed that, for an initial coherent Gibbs state, by neglecting subdominant terms for the symmetric value $q=1/2$ we get a Gaussian probability distribution of work, and so a non-contextual protocol. However, for $q\neq 1/2$, the quasiprobability of work can take negative values depending on the initial state. In contrast, for a local quench there are initial states such that any quasiprobability representation in the class is contextual as signaled by a negative fourth moment. 
We note that the quasiprobability distribution can be measured experimentally in different ways~\cite{Francica22,Francica222}, also by using a qubit (see Appendix~\ref{app. meas}).
In the end, beyond the fundamental purposes of the paper, it is interesting to understand if the contextuality can be related to some advantages from a thermodynamic point of view, however further investigations are needed to going in this direction.
In particular, although the protocol tends to be non-contextual in the thermodynamic limit for a global quench, the initial quantum coherence can be still a useful resource for the work extraction in the protocol when it is correlated with the work.

\subsection*{Acknowledgements}
The authors acknowledge financial support from the project BIRD 2021 "Correlations, dynamics and topology in long-range quantum systems" of the Department of Physics and Astronomy, University of Padova and from the EuropeanUnion-NextGenerationEU within the National Center for HPC, Big Data and Quantum Computing (Project No. CN00000013, CN1 Spoke 10 Quantum Computing).

\appendix
\section{Work moments}\label{app. w moments}
Let us derive a closed formula for the work moments. We define $H=H(\lambda_0)$ and $H'=H^{(H)}(\lambda_\tau)$. The nth work moment can be calculated as
\begin{equation}
\langle w^n \rangle = (-i)^n\partial_u^n\chi_q(0) = \frac{(-i)^n\partial_u^n X_q(0)}{2}+\frac{(-i)^n\partial_u^nX_{1-q}(0)}{2}
\end{equation}
To calculate $(-i)^n\partial_u^n X_q(0)$, we note that
\begin{equation}
X_q(u) = \Tr{\rho_0(u)e^{iuH'}}
\end{equation}
where we have defined
\begin{equation}
\rho_0(u) = e^{-iuqH}\rho_0e^{-iu(1-q)H}
\end{equation}
Then
\begin{equation}
(-i)^n\partial_u^n X_q(u) = \sum_{k=0}^n \binom{n}{k}\Tr{((-i)^{n-k}\partial_u^{n-k}\rho_0(u))H'^k e^{iuH'}}
\end{equation}
where we have noted that $(-i)^k\partial_u^ke^{iuH'} = H'^k e^{iuH'}$. It is easy to see that
\begin{equation}
(-i)^{n}\partial_u^{n}\rho_0(u)= (-1)^n\sum_{k=0}^n\binom{n}{k}(qH)^{n-k}\rho_0(u) ((1-q)H)^k
\end{equation}
from which
\begin{eqnarray}
\nonumber(-i)^n\partial_u^n X_q(0) &=& \sum_{k=0}^n (-1)^{n-k}\binom{n}{k}\sum_{l=0}^{n-k}\binom{n-k}{l}q^{n-k-l}(1-q)^l\\
&&\times\Tr{H^{n-k-l}\rho_0 H^lH'^k }
\end{eqnarray}

\section{Quasiprobability of work}\label{app. chara}
We consider two different initial states, a Gibbs state $\rho_G = e^{-\beta H(\lambda_0)}/Z$, and a coherent Gibbs state $\ket{\Psi_G}$.
In particular, for $\phi_j=0$, the state $\ket{\Psi^s_G}$ in Eq.~\eqref{eq. psi^s_G} reads
\begin{equation}
\ket{\Psi^s_G} = \frac{1}{\sqrt{Z}} \otimes_{k\in K_s} \left( e^{\frac{\beta \epsilon_k}{4}}\ket{\tilde 0_k}+e^{-\frac{\beta \epsilon_k}{4}}\ket{\tilde 1_k}\right)
\end{equation}
It can be expressed as
\begin{eqnarray}
\ket{\Psi^+_G}&=&\frac{1}{\sqrt{Z}}\left(\otimes_{k>0} \ket{\Psi_k}\right)\\
\ket{\Psi^-_G}&=&\frac{1}{\sqrt{Z}}\left(\otimes_{k>0} \ket{\Psi_k}\right)\otimes\ket{\Psi_0}\otimes\ket{\Psi_\pi}
\end{eqnarray}
where $\ket{\Psi_k}=\left(\ket{\tilde 0_k} + e^{- \frac{\beta \epsilon_k}{2}}\ket{\tilde 1_k}\right) \otimes\left(e^{\frac{\beta \epsilon_k}{2}}\ket{\tilde 0_{-k}} + \ket{\tilde 1_{-k}}\right)$.
Thus, by noting that $P_s = (I+se^{i\pi N})/2$, and $e^{i\pi N} = \bra{\tilde 0_s}e^{i\pi N} \ket{\tilde 0_s} e^{i\pi \sum_{k\in K_s}\alpha^\dagger_k \alpha_k}$, we get
\begin{eqnarray}\label{eq. X_q IS}
\nonumber  X_q(u) &=& \frac{1}{2}\sum_s \text{Tr}\big\{e^{-iuqH_s(\lambda_0)}\rho^s_0 e^{-iu(1-q)H_s(\lambda_0)} e^{iu H_s^{(H)}(\lambda_\tau)}\big\}\\
\nonumber && + \eta_s\text{Tr}\big\{  e^{-iuqH_s(\lambda_0)}e^{i\pi \sum_{k\in K_s}\alpha^\dagger_k \alpha_k}\rho^s_0 e^{-iu(1-q)H_s(\lambda_0)} \\
 &&\times e^{iu H_s^{(H)}(\lambda_\tau)}\big\}
\end{eqnarray}
where we have defined $\eta_s = s \bra{\tilde 0_s}e^{i\pi N} \ket{\tilde 0_s}$. Let us focus on the first term in the sum over $s$, which is
\begin{equation}
X^s_q(u)= \text{Tr}\big\{e^{-iuqH_s(\lambda_0)}\rho^s_0 e^{-iu(1-q)H_s(\lambda_0)} e^{iu H_s^{(H)}(\lambda_\tau)}\big\}
\end{equation}
Then, e.g., for $s=-$, to evaluate the trace we can consider the basis formed by the vectors $\ket{\{n_k\}}=(\otimes_{k>0} \ket{n_k n_{-k}})\otimes \ket{n_0}\otimes \ket{n_\pi}$, with $n_k=0,1$, where $\ket{n_k n_{-k}}=(a^\dagger_k)^{n_k}(a^\dagger_{-k})^{n_{-k}}\ket{0_k 0_{-k}}$, where $\ket{0_k}$ is the vacuum state for the fermion $a_k$.
Of course $\{\ket{n_k n_{-k}}\}$ generates an invariant dynamically subspace, and in this subspace the Hamiltonian $H_s(\lambda)$ is the matrix $H_k(\lambda)$ such that
\begin{eqnarray}
H_k(\lambda) \ket{0_k 0_{-k}} &=& 2(\lambda+\cos k)\ket{0_k 0_{-k}} -2i \sin k \ket{1_k 1_{-k}}\\
H_k(\lambda) \ket{1_k 1_{-k}} &=& -2(\lambda+\cos k)\ket{1_k 1_{-k}} +2i \sin k \ket{0_k 0_{-k}}\\
H_k(\lambda) \ket{0_k 1_{-k}} &=& 0\\
H_k(\lambda) \ket{1_k 0_{-k}} &=& 0
\end{eqnarray}
However, it is convenient to consider the initial eigenstates $\ket{\tilde n_k \tilde n_{-k}}$ such that
\begin{equation}
H_k(\lambda_0) \ket{\tilde n_k \tilde n_{-k}} = (\epsilon_k n_k + \epsilon_k (n_{-k}-1)) \ket{\tilde n_k \tilde n_{-k}}
\end{equation}
For our two initial states it is equal to
\begin{equation}
X^s_q(u) = \frac{1}{Z}\prod_{k\in K_s : k\geq 0} X^{(k)}_q(u)
\end{equation}
For the Gibbs state, for $k> 0$ and $k\neq \pi$, we have
\begin{eqnarray}
\nonumber X^{(k)}_q(u) &=& \sum_{n_k,n_{-k}} e^{(-iu-\beta)(\epsilon_k n_k + \epsilon_k (n_{-k}-1))}\\
 &&\times \bra{\tilde n_k \tilde n_{-k}} U_{\tau,0}^\dagger e^{iu H_k(\lambda_\tau)} U_{\tau,0} \ket{\tilde n_k \tilde n_{-k}}
\end{eqnarray}
To evaluate $X^{(k)}_q(u)$, we note that
\begin{equation}
e^{iuH_k(\lambda_\tau)} = e^{-i u \epsilon'_k \hat{d}'_k\cdot \vec\tau} = (\cos(u\epsilon'_k) I - i \sin(u\epsilon'_k) \hat{d}'_k\cdot \vec\tau)\oplus I
\end{equation}
where $\hat{d}'_k=\hat{d}_k(\lambda_\tau)$, $\epsilon'_k=\epsilon_k(\lambda_\tau)$, $\vec{\tau}=(\tau_1,\tau_2,\tau_3)^T$, where $\tau_i$ are the Pauli matrices, i.e., $\tau_3=\ket{0_k 0_{-k}}\bra{0_k 0_{-k}}-\ket{1_k 1_{-k}}\bra{1_k 1_{-k}}$, and so on.
We have to calculate
\begin{eqnarray}
\nonumber &&\bra{\tilde n_k \tilde n_{-k}} U_{\tau,0}^\dagger e^{iu H_k(\lambda_\tau)} U_{\tau,0} \ket{\tilde n_k \tilde n_{-k}}= \cos(u\epsilon'_k) \\
 &&- i  \sin(u\epsilon'_k)\bra{\tilde n_k \tilde n_{-k}} U_{\tau,0}^\dagger \hat{d}'_k\cdot \vec\tau U_{\tau,0} \ket{\tilde n_k \tilde n_{-k}}
\end{eqnarray}
with $(n_k,n_{-k})=(0,0)$ and $(n_k,n_{-k})=(1,1)$, while $\bra{\tilde 0_k \tilde 1_{-k}} U^\dagger_{\tau,0} e^{iu H_k(\lambda_\tau)} U_{\tau,0} \ket{\tilde 0_k \tilde 1_{-k}} = \bra{\tilde 1_k \tilde 0_{-k}} U_{\tau,0}^\dagger e^{iu H_k(\lambda_\tau)} U_{\tau,0} \ket{\tilde1_k \tilde 0_{-k}} = 1$. In particular, since $\hat{d}'_k\cdot \vec\tau$ is traceless, we get $\bra{\tilde 0_k \tilde 0_{-k}} U^\dagger_{\tau,0} \hat{d}'_k\cdot \vec\tau U_{\tau,0} \ket{\tilde 0_k \tilde 0_{-k}} + \bra{\tilde 1_k \tilde 1_{-k}} U_{\tau,0}^\dagger \hat{d}'_k\cdot \vec\tau U_{\tau,0} \ket{\tilde 1_k \tilde 1_{-k}}=0$, from which we get $X^{(k)}_q(u)=X^{(k),th}_q(u)$ with
\begin{eqnarray}
\nonumber && X^{(k),th}_q(u) = 2 \bigg( \cos((u-i\beta)\epsilon_k) \cos(u\epsilon'_k) + \sin((u-i\beta)\epsilon_k) \\
&&\times\sin(u\epsilon'_k)\bra{\tilde 0_k \tilde 0_{-k}} U^\dagger_{\tau,0} \hat{d}'_k\cdot \vec\tau U_{\tau,0} \ket{\tilde 0_k \tilde 0_{-k}} )+1\bigg)
\end{eqnarray}
In contrast, for the coherent Gibbs state, for $k>0$ and $k\neq \pi$ we get
\begin{equation}\label{eq. X_q psi}
X^{(k)}_q(u) = \bra{\Psi_k(q-1)} U_{\tau,0}^\dagger e^{iu H_k(\lambda_\tau)} U_{\tau,0} \ket{\Psi_k(q)}
\end{equation}
where
\begin{equation}
\ket{\Psi_k(q)} = \left(\ket{\tilde 0_k} + e^{-i u q \epsilon_k - \frac{\beta \epsilon_k}{2}}\ket{\tilde 1_k}\right) \otimes\left(e^{i u q \epsilon_k + \frac{\beta \epsilon_k}{2}}\ket{\tilde 0_{-k}} + \ket{\tilde 1_{-k}}\right)
\end{equation}
Thus, we get
\begin{eqnarray}
\nonumber X^{(k)}_q(u) &=& 2\bigg(\cos((u-i\beta)\epsilon_k)\cos(u\epsilon'_k) -\frac{i}{2}\sin(u\epsilon'_k) \\
&&\times\bra{\tilde \Psi_k(q-1)}U_{\tau,0}^\dagger\hat{d}'_k\cdot \vec{\tau} U_{\tau,0}\ket{\tilde \Psi_k(q)} +1\bigg)
\end{eqnarray}
where $\ket{\tilde \Psi_k(q)}= e^{iuq\epsilon_k + \beta \epsilon_k/2}\ket{\tilde 0_k \tilde 0_{-k}}+e^{-iuq\epsilon_k - \beta \epsilon_k/2}\ket{\tilde 1_k \tilde 1_{-k}}$. We get
\begin{equation}
X^{(k)}_q(u) = X^{(k),th}_q(u) +X^{(k),coh}_q(u)
\end{equation}
where the coherent contribution is
\begin{eqnarray}
\nonumber X^{(k),coh}_q(u) &=& - 2i\sin(u\epsilon'_k)\text{Re}\big(e^{-iu(2q-1)\epsilon_k}\\
&&\times\bra{\tilde 0_k\tilde0_{-k}} U_{\tau,0}^\dagger\hat{d}'_k\cdot \vec{\tau} U_{\tau,0} \ket{\tilde 1_k \tilde 1_{-k}}\big)
\end{eqnarray}
To calculate the second term in the sum over $s$ in Eq.~\eqref{eq. X_q IS}, we note that
\begin{eqnarray}
e^{i\pi \sum_{k\in K_s}\alpha^\dagger_k \alpha_k} &=& (-1)^{\frac{L}{2}} e^{i\pi \sum_{k\in K_s}\left(\alpha^\dagger_k \alpha_k-\frac{1}{2}\right)}\\
\nonumber &=& (-1)^{\frac{L}{2}} e^{iq\pi \sum_{k\in K_s}\left(\alpha^\dagger_k \alpha_k-\frac{1}{2}\right)} \\
&&\times e^{i(1-q)\pi \sum_{k\in K_s}\left(\alpha^\dagger_k \alpha_k-\frac{1}{2}\right)}
\end{eqnarray}
then the second term is $\eta_s X'^s_q(u)$, where $ X'^s_q(u)$ is obtained by multiplying $X^s_q(u)$ by $(-1)^{\frac{L}{2}}$ and by performing the substitution $u\epsilon_k\mapsto u\epsilon_k-\pi$ so that
\begin{equation}
X'^s_q(u) = \frac{1}{Z}\prod_{k\in K_s : k\geq 0} X'^{(k)}_q(u)
\end{equation}
with
\begin{equation}
X'^{(k)}_q(u)=  X^{(k)}_q(u)-4
\end{equation}
for $k>0$ and $k\neq \pi$. Then, we get
\begin{equation}
X_q(u) = \frac{1}{2}\sum_s X^s_q(u) + \eta_s X'^s_q(u)
\end{equation}
The partition function can be calculated as
\begin{eqnarray}
Z &=& \Tr{e^{-\beta H(\lambda_0)}} = \sum_s \Tr{P_s e^{-\beta H_s(\lambda_0)}}\\
\nonumber &=& \frac{1}{2}\sum_s  \Tr{e^{-\beta H_s(\lambda_0)}} \\
&&+ \eta_s \Tr{e^{-\beta H_s(\lambda_0)+i\pi\sum_{k\in K_s}\alpha_k^\dagger \alpha_k}}\\
\nonumber &=& \frac{1}{2}\sum_s \prod_{k\in K_s} 2\cosh(\beta \epsilon_k/2) \\
&&+ \eta_s \prod_{k\in K_s} 2\sinh(\beta \epsilon_k/2)
\end{eqnarray}

Concerning the quasiprobability distribution of work $p_q(w)$, it can be calculated from the characteristic function as
\begin{eqnarray}
p_q(w) &=& \int  \frac{e^{-iu w}}{2\pi} \chi_q(u)du \\
&=& \frac{1}{2}\int  \frac{e^{-iu w}}{2\pi}\left(X_q(u)+X_{1-q}(u)\right)du
\end{eqnarray}
Let us focus on the thermodynamic limit. For $|\lambda_0|>1$, we get $Z\sim \prod_{k\in K_+} Z_k$ with $Z_k=2\cosh(\beta \epsilon_k/2)$, and $X_q(u) \sim  X^+_q(u)$.
Then $X_q(u)$ is the product of the characteristic functions having quasiprobability distributions
\begin{equation}
p^{(k)}_q(w) = \frac{1}{Z_k^2} \int  \frac{e^{-iu w}}{2\pi} X^{(k)}_q(u)du
\end{equation}
thus the quasiprobability distribution of work reads
\begin{eqnarray}\label{eq. p para}
\nonumber p_q(w) &=& \frac{1}{2}\int \left(\prod_{k>0}p^{(k)}_q(w_k) + \prod_{k>0}p^{(k)}_{1-q}(w_k)\right) \\
&&\times\delta\left(w-\sum_{k>0} w_k\right) \prod_{k> 0} dw_k
\end{eqnarray}
We note that the average work can be calculated as
\begin{equation}\label{eq. w1}
\langle w \rangle = -i \partial_u \chi_q(0) = -i\sum_{k >0} \frac{1}{Z_k^2} \partial_u  X^{(k)}_q(0)
\end{equation}
On the other hand, for $|\lambda_0|<1$, we get $Z\sim \prod_{k\in K_+} Z_k + \prod_{k\in K_+} Z'_k$ with $Z'_k=2\sinh(\beta \epsilon_k/2)$, and $X_q(u) \sim  X^+_q(u)+X'^+_q(u)$ from which
\begin{equation}
X_q(u) = \gamma \prod_{k>0} \frac{ X^{(k)}_q(u)}{Z_k^2} + (1-\gamma) \prod_{k>0} \frac{ X'^{(k)}_q(u)}{Z'^2_k}
\end{equation}
where $\gamma = (\prod_{k>0} Z_k^2 )/Z $.
In the thermodynamic limit, we get
\begin{equation}
\gamma = \frac{e^{\frac{2L}{\pi}\int_0^\pi \cosh^2(\beta \epsilon_k/2) dk }}{e^{\frac{2L}{\pi}\int_0^\pi \cosh^2(\beta \epsilon_k/2) dk }+e^{\frac{2L}{\pi}\int_0^\pi \sinh^2(\beta \epsilon_k/2) dk }}\to 1
\end{equation}
for a non-zero temperature, since $\int_0^\pi \cosh^2(\beta \epsilon_k/2) dk>\int_0^\pi \sinh^2(\beta \epsilon_k/2) dk $. In contrast, for $\beta\to \infty$, we get $Z'_k\sim Z_k$ so that $\gamma=1/2$, and $X'^{(k)}_q(u)\sim X^{(k)}_q(u)$. Then we get the same expression of the quasiprobability of work of Eq.~\eqref{eq. p para}.

\subsection{Sudden quench}
Let us consider a sudden quench, i.e., the limit $\tau\to 0$, so that $U_{\tau,0}=I$.
For the Gibbs state we get $X^{(k)}_q(u)=X^{(k),th}_q(u)$ with
\begin{eqnarray}\label{eq. X th}
\nonumber X^{(k),th}_q(u) &=& 2 \bigg( \cos((u-i\beta)\epsilon_k) \cos(u\epsilon'_k) + \sin((u-i\beta)\epsilon_k)\\
&& \times\sin(u\epsilon'_k)\hat{d}_k\cdot \hat{d}'_k +1\bigg)
\end{eqnarray}
by noting that
\begin{eqnarray}
\nonumber && \bra{\tilde 0_k\tilde0_{-k}} \hat{d}'_k\cdot \vec\tau  \ket{\tilde 0_k \tilde 0_{-k}} = \Tr{\hat{d}'_k\cdot \vec\tau  \ket{\tilde 0_k \tilde 0_{-k}}\bra{\tilde 0_k\tilde0_{-k}} } \\
&& = \frac{1}{2}\Tr{\hat{d}'_k\cdot \vec\tau  \hat{d}_k\cdot \vec\tau } = \hat{d}_k\cdot \hat{d}'_k
\end{eqnarray}
since $\hat{d}_k\cdot \vec\tau = \ket{\tilde 0_k \tilde 0_{-k}}\bra{\tilde 0_k\tilde0_{-k}}-\ket{\tilde 1_k \tilde 1_{-k}}\bra{\tilde 1_k\tilde1_{-k}}$.
On the other hand, for the coherent Gibbs state we get
\begin{equation}
X^{(k)}_q(u) = X^{(k),th}_q(u) +X^{(k),coh}_q(u)
\end{equation}
To evaluate the coherent contribution $X^{(k),coh}_q(u)$, we note that
\begin{eqnarray}
\nonumber &&\bra{\tilde 0_k\tilde0_{-k}} \hat{d}'_k\cdot \vec\tau  \ket{\tilde 1_k \tilde 1_{-k}} = \Tr{\hat{d}_k\cdot \vec\tau \hat{d}'_k\cdot \vec\tau  \ket{\tilde 1_k \tilde 1_{-k}}\bra{\tilde 0_k\tilde0_{-k}} }\\
 &&= i\Tr{(\hat{d}_k\times \hat{d}'_k )\cdot \vec \tau \ket{\tilde 1_k \tilde 1_{-k}}\bra{\tilde 0_k\tilde0_{-k}} }
\end{eqnarray}
then
\begin{equation}
\bra{\tilde 0_k\tilde0_{-k}} \hat{d}'_k\cdot \vec\tau  \ket{\tilde 1_k \tilde 1_{-k}}  = i(\hat{d}_k\times \hat{d}'_k)_x \bra{\tilde 0_k\tilde0_{-k}} \tau_1 \ket{\tilde 1_k \tilde 1_{-k}}
\end{equation}
since $\hat{d}_k\times \hat{d}'_k$ has only x-component. Since $\bra{\tilde 0_k\tilde0_{-k}} \tau_1  \ket{\tilde 1_k \tilde 1_{-k}}=1$, we get
\begin{equation}
\bra{\tilde 0_k\tilde0_{-k}} \hat{d}'_k\cdot \vec\tau  \ket{\tilde 1_k \tilde 1_{-k}}  = i(\hat{d}_k\times \hat{d}'_k)_x
\end{equation}
Thus we get
\begin{equation}
X^{(k),coh}_q(u) = - 2i\sin(u\epsilon'_k)\sin(u(2q-1)\epsilon_k)(\hat{d}_k\times \hat{d}'_k)_x
\end{equation}
For $\phi_j\neq 0$, we have  $\ket{\Psi_k}=\left(\ket{\tilde 0_k} + e^{i\phi_k- \frac{\beta \epsilon_k}{2}}\ket{\tilde 1_k}\right) \otimes\left(e^{\frac{\beta \epsilon_k}{2}}\ket{\tilde 0_{-k}} + e^{i\phi_{-k}}\ket{\tilde 1_{-k}}\right)$, with $\phi_{-k}=\phi_{k}$. Thus, by considering the corresponding state $\ket{\Psi_k(q)}$, the only effect of the phase $\phi_k$ is the shift $uq\epsilon_k \to uq\epsilon_k -\phi_k $, then we get
\begin{equation}
X^{(k),coh}_q(u) = - 2i\sin(u\epsilon'_k)\sin(u(2q-1)\epsilon_k-2\phi_k)(\hat{d}_k\times \hat{d}'_k)_x
\end{equation}

\subsection{Arbitrary quench}
In the end, let us consider an arbitrary quench. The time evolution acts as a rotation of  the vector $\vec{d}_k(\lambda_\tau)$, so that $U_{\tau,0}^\dagger \hat{d}_k (\lambda_\tau)\cdot \vec{\tau} U_{\tau,0} = \hat{d}'_k(\lambda_\tau)\cdot \vec{\tau} $, where for brevity $\hat{d}'_k=\hat{d}'_k(\lambda_\tau)$. Then, $ X^{(k),th}_q(u)$ is still given by Eq.~\eqref{eq. X th} with the new vector $\hat{d}'_k$, and if $\hat{d}_k\times \hat{d}'_k$ has also a non-zero y-component, then
\begin{equation}
\bra{\tilde 0_k\tilde0_{-k}} \hat{d}'_k\cdot \vec\tau  \ket{\tilde 1_k \tilde 1_{-k}}  = i(\hat{d}_k\times \hat{d}'_k)_x+(\hat{d}_k\times \hat{d}'_k)_y
\end{equation}
from which the coherence contribution has a further term and reads
\begin{eqnarray}\label{eq. app arb quench}
\nonumber X^{(k),coh}_q(u) &=& - 2i\sin(u\epsilon'_k) \big( \sin(u(2q-1)\epsilon_k-2\phi_k)(\hat{d}_k\times \hat{d}'_k)_x\\
&& + \cos(u(2q-1)\epsilon_k-2\phi_k)(\hat{d}_k\times \hat{d}'_k)_y \big)
\end{eqnarray}

\subsection{Histogram}
To determinate the quasiprobability distribution of work from the characteristic function $\chi_q(u)$ we consider the intervals $I_n = [w_n-\Delta w/2,w_n + \Delta w/2]$, where $w_n=n \Delta w$ with $n$ integer. Then, we can determinate the histogram by calculating the probability
\begin{equation}
p_n = \int_{I_n} p_q(w) dw = \frac{\Delta w}{2\pi}\int  \chi_q(u)\text{sinc}\left(\frac{u\Delta w}{2}\right) e^{-i u w_n} du
\end{equation}
where $\text{sinc}(x) = \sin(x)/x$. To calculate the integral we can focus on the interval $[-2\pi K /\Delta w,2\pi K /\Delta w]$ with $K$ large enough. Of course $p_q(w_n)\approx p_n/\Delta w$ for $\Delta w$ small enough.

\section{Superposition of two coherent Gibbs states}\label{app. super}
For simplicity we consider the fermionic Hamiltonian $H_+$. We focus on the initial state
\begin{equation}
\ket{\Psi} = a \ket{\Psi^+_{G,1}} + b \ket{\Psi^+_{G,2}}
\end{equation}
where $\ket{\Psi^+_{G,i}}$ is the coherent Gibbs state
\begin{equation}
\ket{\Psi^+_{G,i}} = \otimes_{k>0} \frac{\ket{\Psi_{i,k}}}{Z_{i,k}}
\end{equation}
where
\begin{equation}
\ket{\Psi_{i,k}}=\left(\ket{\tilde 0_k} + e^{i\phi_{i,k}- \frac{\beta_i \epsilon_k}{2}}\ket{\tilde 1_k}\right) \otimes\left(e^{\frac{\beta_i \epsilon_k}{2}}\ket{\tilde 0_{-k}} + e^{i\phi_{i,-k}}\ket{\tilde 1_{-k}}\right)
\end{equation}
We will get
\begin{eqnarray}
\nonumber X^+_q(u) &=& |a|^2 \prod_{k>0}  \frac{X^{(k)}_{q,1}(u)}{Z_{1,k}^2}+|b|^2 \prod_{k>0}  \frac{X^{(k)}_{q,2}(u)}{Z_{2,k}^2} \\
 &&+ ab^* \prod_{k>0}  \frac{Y^{(k)}_{q,1}(u)}{Z_{1,k}Z_{2,k}}+ a^*b \prod_{k>0}  \frac{Y^{(k)}_{q,2}(u)}{Z_{1,k}Z_{2,k}}
\end{eqnarray}
where
\begin{eqnarray}
X^{(k)}_{q,i}(u) &=& \bra{\Psi_{i,k}(q-1)} U_{\tau,0}^\dagger e^{iu H_k(\lambda_\tau)} U_{\tau,0} \ket{\Psi_{i,k}(q)}\\
Y^{(k)}_{q,1}(u) &=& \bra{\Psi_{2,k}(q-1)} U_{\tau,0}^\dagger e^{iu H_k(\lambda_\tau)} U_{\tau,0} \ket{\Psi_{1,k}(q)}\\
Y^{(k)}_{q,2}(u) &=& \bra{\Psi_{1,k}(q-1)} U_{\tau,0}^\dagger e^{iu H_k(\lambda_\tau)} U_{\tau,0} \ket{\Psi_{2,k}(q)}
\end{eqnarray}
with
\begin{equation}
\ket{\Psi_{i,k}}=\left(\ket{\tilde 0_k} + e^{-i u q \epsilon_k+i\phi_{i,k}- \frac{\beta_i \epsilon_k}{2}}\ket{\tilde 1_k}\right) \otimes\left(e^{i u q \epsilon_k+\frac{\beta_i \epsilon_k}{2}}\ket{\tilde 0_{-k}} + e^{i\phi_{i,-k}}\ket{\tilde 1_{-k}}\right)
\end{equation}
Then we can write
\begin{equation}
X^+_q(u) = |a|^2 e^{L g_{1,q}(u)}+|b|^2 e^{L g_{2,q}(u)} + a b^* e^{L y_{1,q}(u)}+ a^* b e^{L y_{2,q}(u)}
\end{equation}
where
\begin{eqnarray}
g_{i,q}(u) &=& \frac{1}{2\pi}\int_0^\pi \ln\left(\frac{X^{(k)}_{q,i}(u)}{Z_{i,k}^2}\right) dk\\
y_{i,q}(u) &=& \frac{1}{2\pi}\int_0^\pi \ln\left(\frac{Y^{(k)}_{q,i}(u)}{Z_{1,k}Z_{2,k}}\right) dk
\end{eqnarray}
As $L\to \infty$, the Fourier transform of $e^{L y_{i,q}(u)}$ gives
\begin{equation}
p'_{i,q}(w) \sim  \frac{e^{L y_{i,q}(0)}}{\sqrt{2\pi x^{(2)}_{i,q}}} e^{-\frac{\left(w-x^{(1)}_{i,q}\right)^2}{2x^{(2)}_{i,q}}}
\end{equation}
where $x^{(1)}_{i,q}=-i \partial_u y_{i,q}(0)L$ and  $x^{(2)}_{i,q}=- \partial^2_u y_{i,q}(0)L$. For $q=1/2$, we get $x^{(i)}_{2,1/2}=\left(x^{(i)}_{1,1/2}\right)^*$, then the quasiprobability distribution of work reads
\begin{eqnarray}
\nonumber p_{1/2}(w) &\sim& |a|^2\frac{e^{-\frac{(w -\bar{w}_1)^2}{2 v_{1,q}}}}{\sqrt{2\pi v_{1,q}}} + |b|^2 \frac{e^{-\frac{(w -\bar{w}_2)^2}{2 v_{2,q}}}}{\sqrt{2\pi v_{2,q}}}\\
 && + 2 \text{Re}\left ( a b^* p'_{1,1/2}(w)\right)
\end{eqnarray}
where $\bar{w}_{i}=-i \partial_u g_{i,q}(0)L$ and  $v_{i,q}=- \partial^2_u g_{i,q}(0)L$, so that $p_{1/2}(w)$ can take negative values.
However, since the real part of $y_{i,q}(0)$ is negative, $p'_{i,q}(w)$ tends exponentially to zero in the thermodynamic limit and $p_{1/2}(w)$ is the convex combination of two Gaussian probability distributions, which is positive.
\subsection{Generalized coherent Gibbs state}\label{app. im part r_q}

For an arbitrary quench from the initial coherent Gibbs state, from Eq.~\eqref{eq. app arb quench}, we get
\begin{eqnarray}
\nonumber r_q &=& \frac{(2 q-1)L}{2\pi}\int_0^\pi \frac{\epsilon_k \epsilon'_k}{\cosh^2(\beta\epsilon_k/2)}(\cos(2\phi_k)(\hat{d}_k\times \hat{d}'_k)_x \\
&&+ \sin(2\phi_k)(\hat{d}_k\times \hat{d}'_k)_y)dk
\end{eqnarray}
which is zero for $q=1/2$.
Let us focus on an initial state of the form
\begin{equation}\label{app. psi two}
\ket{\Psi^+} = \otimes_{k>0} \ket{\Psi_k}
\end{equation}
which generalizes the coherent Gibbs state $\ket{\Psi^+_G}$, where we have defined the states
\begin{equation}
\ket{\Psi_k} = \sum_{n_k,n_{-k}} c_{n_k n_{-k}} \ket{\tilde n_k\tilde n_{-k}}
\end{equation}
This implies that $X_q(u)$ has the form in Eq.~\eqref{eq. XQQ} with $Z_k=1$,
where $X^{(k)}_q(u)$ can be calculated from Eq.~\eqref{eq. X_q psi} with
\begin{eqnarray}
\nonumber \ket{\Psi_k(q)} &=& c_{00} e^{iuq\epsilon_k}\ket{\tilde 0_k\tilde 0_{-k}} + c_{11} e^{-iuq\epsilon_k}\ket{\tilde 1_k\tilde 1_{-k}}\\
 &&+ c_{01} \ket{\tilde 0_k \tilde 1_{-k}}+ c_{10} \ket{\tilde 1_k\tilde 0_{-k}}
\end{eqnarray}
Then, the representation for $q=1/2$ will be non-contextual. Let us show explicitly that $r_{1/2}=0$.
$X^{(k)}_q(u)$ reads
\begin{equation}
X^{(k)}_q(u) = X^{(k)}_{no}(u) + \delta X^{(k)}_{q}(u)
\end{equation}
where $X^{(k)}_{no}(u)$ does not depend on $q$, and
\begin{eqnarray}
\nonumber \delta X^{(k)}_q(u)  &=& -2i\sin(u\epsilon'_k)\text{Re}\bigg( c_{00}^*c_{11} e^{-iu(2q-1)\epsilon_k}  \big(i(\hat{d}_k\times \hat{d}'_k)_x\\
&&+(\hat{d}_k\times \hat{d}'_k)_y\big) \bigg)
\end{eqnarray}
Then, $\partial^2_u \delta X^{(k)}_q(0)$  is imaginary and $\partial^2_u \delta X^{(k)}_q(0) \propto (1-2q)$. Similarly, it is easy to see that $\partial^2_u X^{(k)}_{no}(0)$ is real. Furthermore, $\partial_u X^{(k)}_q(0)$  is imaginary, then $r_q$ is obtained by calculating an integral with respect to $k$ of $\partial^2_u \delta X^{(k)}_q(0)$, so that we get $r_q \propto (1-2q)$, which is zero for $q=1/2$.
In the end, we note that a linear combination of states of the form in Eq.~\eqref{app. psi two} will give for $q=1/2$ a convex combination of Gaussian probability distributions, which is positive.

\section{Negativity}\label{app. negativity}
To prove that $\mathcal N=1$ implies in general that $p_q(w)\geq 0$, we can proceed ad absurdum. We write $p_q(w) = p(w) + \delta p(w)$ where $p(w)\geq 0$, $\int p(w) dw=1$ and $\int \delta p(w)dw =0$. If $p_q(w)\geq 0$ for $w\in I$ and $p_q(w)<0$ for $w\in I'$, then $\delta p(w) < 0 $  for $w\in I'$ and $I=I_+\cup I_-$ such that $\delta p(w) \geq 0 $  for $w\in I_+$ and $\delta p(w) < 0 $  for $w\in I_-$. Then, from $\mathcal N =1$, we get the condition $p(I)-p(I')+\delta p(I_+) + \delta p(I_-)-\delta p(I')=1$, where $p(I)=\int_I p(w) dw$ and so on, thus we get the system
\begin{equation}
    \begin{cases}
      p(I)+p(I')=1\\
      p(I)\geq 0\\
      p(I')\geq 0\\
      \delta p(I_+) + \delta p(I-) + \delta p(I') =0\\
      \delta p(I_+)\geq 0\\
      \delta p(I_-)<0\\
      \delta p(I')<0\\
      p(I)-p(I')+\delta p(I_+) + \delta p(I_-)-\delta p(I')=1\\
    \end{cases}
\end{equation}
which admits as solution $p(w)$ such that $0\leq p(I)<1$ and $p(I')=1-p(I)$ and $\delta p(w)$ such that $\delta p (I_+) > (1-p(I)+p(I'))/2$, $\delta p(I_-)=(1-2\delta p(I_+)-p(I)+p(I'))/2$ and $\delta p(I')=-\delta p(I_-)-\delta p(I_+)$. Then $\delta p(I') = -p(I')$, so that $p_q(I')=0$, which implies that $p_q(w)$ is non-negative.

\section{General quadratic form in Fermi operators}\label{app. quadratic form}
We consider the initial Hamiltonian
\begin{equation}
H = \sum_{i,j}\left( a_i^\dagger A_{ij} a_j + \frac{1}{2}\left( a_i^\dagger B_{ij}a_j^\dagger + H.c. \right)\right) - \frac{1}{2}\sum_iA_{ii}
\end{equation}
where $A$ and $B$ are real matrices such that $A^T=A$ and $B^T=-B$. The Hamiltonian can be diagonalized by performing the transformation
\begin{equation}
\alpha_k = \sum_i g_{ki} a_i + h_{ki}a^\dagger_i
\end{equation}
so that
\begin{equation}
H = \sum_k \epsilon_k \left(\alpha^\dagger_k\alpha_k -\frac{1}{2}\right)
\end{equation}
In detail the matrices $g$ and $h$ are such that  $\phi=g+h$ and $\psi= g-h$, where  $\phi$ and $\psi$ are orthogonal matrices such that $\psi^T \epsilon \phi = A+B$, where $\epsilon$ is the diagonal matrix with entries $\epsilon_k$. The final time-evolved Hamiltonian is $H'$ with matrices $A'$ and $B'$, and will be diagonalized by performing the transformation
\begin{equation}
\alpha'_k = \sum_i g'_{ki} a_i + h'_{ki}a^\dagger_i
\end{equation}
so that
\begin{equation}
H' = \sum_k \epsilon'_k \left(\alpha'^\dagger_k\alpha'_k -\frac{1}{2}\right)
\end{equation}

Let us proceed with our investigation by considering the initial state
\begin{equation}\label{eq. psi p 1}
\ket{\Psi_1} = \frac{e^{\frac{\beta}{4}\sum_k\epsilon_k}}{\sqrt{Z_1}}\exp\left(\sum_ke^{-\frac{\beta \epsilon_k}{2}}\alpha_k^\dagger\right)\ket{\tilde 0}
\end{equation}
We note that for $\beta\to\infty$ we get $Z_1\sim Z = \prod_k 2 \cosh(\beta \epsilon_k/2)$ and $\ket{\Psi_1}\sim \ket{\Psi_G}$. We aim to calculate
\begin{equation}
X_q(u) = \bra{\Psi_1} e^{-iu(1-q) H} e^{iu H'} e^{-iuqH} \ket{\Psi_1}
\end{equation}
We consider the vacuum state $\ket{\tilde 0'}$ of the fermions $\alpha'_k$, we get the relation
\begin{equation}
\ket{\tilde 0} = K e^{\frac{1}{2}\sum_{k,k'} G_{kk'}\alpha'^\dagger_k\alpha'^\dagger_{k'}}\ket{\tilde 0'}
\end{equation}
where $G$ is solution of the equation $\tilde g G + \tilde h =0$, where $\tilde g = g g'^T+h h'^T$ and $\tilde h = g h'^T+h g'^T$. In particular,
\begin{equation}
\alpha_k = \sum_{k'} \tilde g_{kk'}\alpha'_{k'} + \tilde h_{kk'}\alpha'^\dagger_{k'}
\end{equation}
We get
\begin{eqnarray}\label{eq. coe 01}
\nonumber &&X_q(u) = |K|^2\frac{e^{(\beta+iu)\sum_k\epsilon_k/2-iu\sum_k\epsilon'_k/2}}{Z_1}\bra{\tilde 0'} \exp\bigg(-\frac{1}{2}\sum_{k,k'} G_{kk'}\\
\nonumber &&\times\alpha'_k \alpha'_{k'}\bigg)\exp\bigg(\sum_k u_k \alpha'_k+v_k \alpha'^\dagger_k\bigg)\exp\bigg(\sum_k u'_k \alpha'^\dagger_k+v'_k \alpha'_k\bigg)\\
&&\times\exp\bigg(\frac{1}{2}\sum_{k,k'} \tilde G_{kk'}\alpha'^\dagger_k \alpha'^\dagger_{k'}\bigg) \ket{\tilde 0'}
\end{eqnarray}
where $\tilde G_{kk'}= G_{kk'}e^{iu(\epsilon'_k+\epsilon'_{k'})}$ and
\begin{eqnarray}
u_k&=& \sum_{k'} e^{-(\beta/2+iu(1-q))\epsilon_{k'}}\tilde g_{k'k}\\
v_k&=& \sum_{k'} e^{-(\beta/2+iu(1-q))\epsilon_{k'}}\tilde h_{k'k}\\
u'_k&=& \sum_{k'} e^{-(\beta/2+iuq)\epsilon_{k'}+iu\epsilon'_k}\tilde g_{k'k}\\
v'_k&=&  \sum_{k'} e^{-(\beta/2+iuq)\epsilon_{k'}-iu\epsilon'_k}\tilde h_{k'k}
\end{eqnarray}
We note that
\begin{eqnarray}
\nonumber &&\exp\bigg(\sum_k u_k \alpha'_k+v_k \alpha'^\dagger_k\bigg)\exp\bigg(\sum_k u'_k \alpha'^\dagger_k+v'_k \alpha'_k\bigg)= 1\\
\nonumber&& +\sum_{k,k'} u_k u'_{k'} \alpha'_k \alpha'^\dagger_{k'}+u_k v'_{k'} \alpha'_k \alpha'_{k'}+v_k u'_{k'} \alpha'^\dagger_k \alpha'^\dagger_{k'}-v'_k v_{k'} \alpha'_k \alpha'^\dagger_{k'}\\
&& + \sum_k v_k v'_k + \cdots
\end{eqnarray}
where we have omitted terms linear in the Fermi operators.
Then, the overlap in Eq.~\eqref{eq. coe 01} can be easily calculated by using the coherent states $\ket{\xi}$ such that $\alpha'_k \ket{\xi} = \xi_k \ket{\xi}$. By using the identity $\int d\xi^* d\xi e^{-\sum_k \xi^*_k\xi_k}\ket{\xi}\bra{\xi}=1$, we get
\begin{eqnarray}
\nonumber && X_q(u) \sim |K|^2\frac{e^{(\beta+iu)\sum_k\epsilon_k/2-iu\sum_k\epsilon'_k/2}}{Z_1}\bigg[ \int d\xi^* d\xi  \\
\nonumber && \times \exp\bigg(-\frac{1}{2}\sum_{k,k'} G_{kk'}\xi_k \xi_{k'}+\sum_{k,k'}( u_k u'_{k'} \xi_k \xi^*_{k'}+u_k v'_{k'} \xi_k \xi_{k'}\\
\nonumber && +v_k u'_{k'} \xi^*_k \xi^*_{k'}-v'_k v_{k'} \xi_k \xi^*_{k'}) -\sum_k \xi^*_k\xi_k+\frac{1}{2}\sum_{k,k'} \tilde G_{kk'}\xi^*_k \xi^*_{k'} \bigg)\\
\nonumber && + \sum_k v_k v'_k\int d\xi^* d\xi  \exp\bigg(-\frac{1}{2}\sum_{k,k'} G_{kk'}-\sum_k \xi^*_k\xi_k\\
&&+\frac{1}{2}\sum_{k,k'} \tilde G_{kk'}\xi^*_k \xi^*_{k'} \bigg) \bigg]
\end{eqnarray}
By performing the integral, we get
\begin{equation}
X_q(u)\sim Ce^{iu\sum_k(\epsilon_k-\epsilon'_k)/2}\bigg( \sqrt{\det(\Gamma(u))}+\sqrt{\det(\Gamma_0(u))}\sum_k  v_k v'_k \bigg)
\end{equation}
where
\begin{equation}\label{eq. gamma_0}
\Gamma_0(u) = \left(
               \begin{array}{cc}
                 G & -I \\
                 I & -\tilde G \\
               \end{array}
             \right)
\end{equation}
and
\begin{equation}
\Gamma(u) = \left(
               \begin{array}{cc}
                 G -M_1& -I-M_2 \\
                 I+M_2^T & -\tilde G-M_3 \\
               \end{array}
             \right)=\Gamma_0(u)+M(u)
\end{equation}
where $M_{1,kk'}=u_kv'_{k'}-u_{k'}v'_{k}$, $M_{2,kk'}=u_ku'_{k'}-v'_kv_{k'}$ and $M_{3,kk'}=v_ku'_{k'}-v_{k'}u'_{k}$. The constant $C$ can be determined by requiring that $X_q(0)=1$. The exact expression of $X_q(u)$ can be obtained by expanding $\sqrt{\det(\Gamma(u))}$ at the first order in $M(u)$, i.e.,
\begin{eqnarray}\label{eq. sec ord}
\nonumber X_q(u)&=&C e^{iu\sum_k(\epsilon_k-\epsilon'_k)/2}\sqrt{\det(\Gamma_0(u))}\bigg( 1+ \frac{1}{2}\Tr{\Gamma_0^{-1}(u) M(u)}\\
&&+\sum_k  v_k v'_k \bigg)
\end{eqnarray}
Concerning the coherent Gibbs state, for low temperatures $\beta \to \infty$ we get
\begin{equation}\label{eq. psi G p}
\ket{\Psi_G} \sim \frac{e^{\frac{\beta }{4}\sum_k\epsilon_k}}{\sqrt{Z}}\left(1+\sum_ke^{-\frac{\beta \epsilon_k}{2}}\alpha_k^\dagger+ \sum_{k>k'}e^{-\frac{\beta (\epsilon_k+\epsilon_{k'})}{2}}\alpha_k^\dagger \alpha_{k'}^\dagger \right)\ket{\tilde 0}
\end{equation}
We define
\begin{eqnarray}
u_{k q }&=&  e^{-(\beta/2+iu(1-q))\epsilon_{k}}\tilde g_{kq}\\
v_{k q}&=&  e^{-(\beta/2+iu(1-q))\epsilon_{k}}\tilde h_{kq}\\
u'_{k q}&=& e^{-(\beta/2+iuq)\epsilon_{k}+iu\epsilon'_q}\tilde g_{kq}\\
v'_{k q} &=&  e^{-(\beta/2+iuq)\epsilon_{k}-iu\epsilon'_q}\tilde h_{kq}
\end{eqnarray}
so that $u_k=\sum_{k'} u_{k' k}$ and so on, then the matrices $V_{1}$, $V_{2}$, $V_{3}$, $V'_{1}$, $V'_{2}$ and $V'_{3}$ with elements $V_{1,qq'} = \sum_{k,k'} s_{k,k'} u_{kq} u_{k'q'}$, $V_{2,qq'} = \sum_{k,k'} s_{k,k'} u_{kq} v_{k'q'}$, $V_{3,qq'} = \sum_{k,k'} s_{k,k'} v_{kq} v_{k'q'}$, $V'_{1,qq'} = \sum_{k,k'} s_{k,k'} v'_{kq} v'_{k'q'}$, $V'_{2,qq'} = \sum_{k,k'} s_{k,k'} v'_{kq} u'_{k'q'}$, $V'_{3,qq'} = \sum_{k,k'} s_{k,k'} u'_{kq} u'_{k'q'}$, where $s_{k,k'}=1$ if $k>k'$,  $s_{k,k'}=-1$ if $k<k'$ and  $s_{k,k}=0$.
Thus, by proceeding similarly, we get at the second order
\begin{eqnarray}\label{eq. sec ord 22}
\nonumber X_q(u)&\sim&C e^{iu\sum_k(\epsilon_k-\epsilon'_k)/2}\sqrt{\det(\Gamma_0(u))}\bigg( 1+ \frac{1}{2}\Tr{\Gamma_0^{-1}(u) M(u)}\\
\nonumber&&+\sum_k  v_k v'_k + \frac{1}{2}\Tr{\Gamma_0^{-1}(u) (V(u)-V'(u))} \\
&&+\frac{1}{2}\Tr{V_{2}-V'_{2}}\bigg)
\end{eqnarray}
where we have defined the matrices
\begin{equation}
V(u) = \left(
               \begin{array}{cc}
                 V_1& V_2 \\
                 -V_2^T & V_3 \\
               \end{array}
             \right)\,,\quad V'(u) = \left(
               \begin{array}{cc}
                 V'_1& V'_2 \\
                 -V'^T_2 & V'_3 \\
               \end{array}
             \right)
\end{equation}
We note that for an initial state that is the ground-state of $H$, we get the characteristic function
\begin{equation}
\chi^{(0)}(u) = e^{iu\sum_k(\epsilon_k-\epsilon'_k)/2}\sqrt{\frac{\det(\Gamma_0(u))}{\det(\Gamma_0(0))}}
\end{equation}
which is obtained from $X_q(u)$ in the limit $\beta\to\infty$. Alternatively, by considering $\theta^T = (\xi^T,{\xi^*}^T)$, Eq.~\eqref{eq. sec ord} can be derived with the help of the identity
\begin{equation}\label{eq. ide 2}
\int d\theta \theta_i \theta_j e^{-\frac{1}{2}\theta^T \Gamma_0 \theta} = -\frac{1}{2} \Tr{\Gamma_0^{-1}X_{ij}}\sqrt{\det(\Gamma_0)}
\end{equation}
where $X_{ij}=\ket{i}\bra{j}-\ket{j}\bra{i}$, and $\ket{i}$ is the unit vector with only the i-th component which is nonzero. Actually $\sqrt{\det(\Gamma_0)}$ is the Pfaffian of $\Gamma_0$. To prove it, we note that
\begin{eqnarray}
\nonumber \int d\theta \theta_i \theta_j e^{-\frac{1}{2}\theta^T \Gamma_0 \theta} &=& \frac{1}{\epsilon}\bigg(\int d\theta (1+\epsilon\theta_i \theta_j) e^{-\frac{1}{2}\theta^T \Gamma_0 \theta} \\
&& - \int d\theta  e^{-\frac{1}{2}\theta^T \Gamma_0 \theta}\bigg)
\end{eqnarray}
The second integral is $\int d\theta  e^{-\frac{1}{2}\theta^T \Gamma_0 \theta}= \sqrt{\det(\Gamma_0)}$. By considering the limit $\epsilon\to 0$, we get
\begin{equation}
\int d\theta \theta_i \theta_j e^{-\frac{1}{2}\theta^T \Gamma_0 \theta} \sim \frac{1}{\epsilon}\left(\int d\theta e^{\frac{\epsilon}{2}(\theta_i \theta_j-\theta_j\theta_i)} e^{-\frac{1}{2}\theta^T \Gamma_0 \theta} - \sqrt{\det(\Gamma_0)}\right)
\end{equation}
then
\begin{equation}
\int d\theta \theta_i \theta_j e^{-\frac{1}{2}\theta^T \Gamma_0 \theta} \sim \frac{1}{\epsilon}\left(\sqrt{\det(\Gamma_0-\epsilon X_{ij})} - \sqrt{\det(\Gamma_0)}\right)
\end{equation}
by evaluating the limit $\epsilon\to 0$, we get Eq.~\eqref{eq. ide 2}. Similarly, we have the identity
\begin{eqnarray}\label{eq. ide 4}
\nonumber &&\int d\theta \theta_i \theta_j \theta_k \theta_l e^{-\frac{1}{2}\theta^T \Gamma_0 \theta} = -\frac{1}{2} \Tr{\Gamma_0^{-1}X_{ij}\Gamma_0^{-1}X_{kl}}\sqrt{\det(\Gamma_0)}\\
&& + \frac{1}{4}\Tr{\Gamma_0^{-1}X_{ij}}\Tr{\Gamma_0^{-1}X_{kl}}\sqrt{\det(\Gamma_0)}
\end{eqnarray}
To prove it, we consider that
\begin{eqnarray}
\nonumber \int d\theta \theta_i \theta_j \theta_k \theta_l e^{-\frac{1}{2}\theta^T \Gamma_0 \theta} &=& \frac{1}{\epsilon}\bigg(\int d\theta \theta_i \theta_j (1+\epsilon\theta_k \theta_l) e^{-\frac{1}{2}\theta^T \Gamma_0 \theta}\\
 && - \int d\theta \theta_i \theta_j  e^{-\frac{1}{2}\theta^T \Gamma_0 \theta}\bigg)
\end{eqnarray}
which, in the limit $\epsilon\to0$ can be evaluated with the help of the identity in Eq.~\eqref{eq. ide 2}. We get
\begin{eqnarray}
\nonumber && \int d\theta \theta_i \theta_j \theta_k \theta_l e^{-\frac{1}{2}\theta^T \Gamma_0 \theta} \sim \frac{1}{2\epsilon}\bigg(\Tr{\Gamma_0^{-1}X_{ij}}\sqrt{\det(\Gamma_0)}\\
&&-\Tr{(\Gamma_0-\epsilon X_{kl})^{-1}X_{ij}}\sqrt{\det(\Gamma_0-\epsilon X_{kl})}\bigg)
\end{eqnarray}
by evaluating the limit $\epsilon\to 0$, we get Eq.~\eqref{eq. ide 4}.
In the end, we consider the initial state in Eq.~\eqref{eq. psi G p}, which is
\begin{equation}\label{eq. psi p}
\ket{\Psi_2} = \frac{e^{\frac{\beta }{4}\sum_k\epsilon_k}}{\sqrt{Z_2}}\left(1+\sum_ke^{-\frac{\beta \epsilon_k}{2}}\alpha_k^\dagger+ \frac{1}{2} \sum_{k,k'}s_{k,k'}e^{-\frac{\beta (\epsilon_k+\epsilon_{k'})}{2}}\alpha_k^\dagger \alpha_{k'}^\dagger \right)\ket{\tilde 0}
\end{equation}
By using the identities in Eqs.~\eqref{eq. ide 2} and~\eqref{eq. ide 4}, we get
\begin{eqnarray}\label{eq. psi pr}
\nonumber X_q(u)&=&C e^{iu\sum_k(\epsilon_k-\epsilon'_k)/2}\sqrt{\det(\Gamma_0(u))}\bigg( 1+ \frac{1}{2}\Tr{\Gamma_0^{-1}(u) M(u)}\\
\nonumber&&+\sum_k  v_k v'_k + \frac{1}{2}\Tr{\Gamma_0^{-1}(u) (V(u)-V'(u))} \\
\nonumber &&+\frac{1}{2}\Tr{V_{2}-V'_{2}}-\frac{1}{4}\Tr{V_2}\Tr{V'_2}\\
\nonumber &&-\frac{1}{4}\Tr{V_2}\Tr{\Gamma^{-1}_0(u)V'(u)}\\
\nonumber &&-\frac{1}{4}\Tr{V'_2}\Tr{\Gamma^{-1}_0(u)V(u)}-\frac{1}{2}\Tr{V_3V'_1}\\
\nonumber && +\frac{1}{2}\Tr{\Gamma^{-1}_0(u)V''(u)}+\frac{1}{2}\Tr{\Gamma^{-1}_0(u)V(u)\Gamma^{-1}_0(u)V'(u)}\\
&& - \frac{1}{4}\Tr{\Gamma^{-1}_0(u)V(u)}\Tr{\Gamma^{-1}_0(u)V'(u)}\bigg)
\end{eqnarray}
where we have defined
\begin{equation}
V''(u) = \left(
               \begin{array}{cc}
                 V_2V'_1+V'_1V^T_2& V_2V'_2-V'_1V_3 \\
                 V_3V'_1-V'^T_2V_2^T & V_3V'_2+V'^T_2V_3 \\
               \end{array}
             \right)
\end{equation}
If we introduce a relative phase $\phi_k$, we have to multiply $u_{k q }$ and $v_{k q }$ by $e^{-i\phi_k}$ and $u'_{k q }$ and $v'_{k q }$ by $e^{i\phi_k}$.

If $A$ and $B$ are complex matrices, we get $g$ and $h$ complex. In this case we have same formulas, with $\tilde g = g g'^\dagger+h h'^T$ and $\tilde h = g h'^T+h g'^\dagger$, and in $\Gamma_0$ in Eq.~\eqref{eq. gamma_0}, we have $G^*$ instead of $G$, and in $u_k$, $v_k$, $u_{kq}$ and $v_{kq}$ we have $\tilde g^*$ and $\tilde h^*$ instead of $\tilde g$ and $\tilde h$.


\section{Initial quantum coherence}\label{app. ini cohe}
We consider the initial state $\rho_0 = \eta \ket{\Psi_G(0)}\bra{\Psi_G(0)} + (1-\eta)\rho_G(0)$, we get
\begin{equation}
\langle e^{it C}\rangle = \Tr{\rho_0 e^{i\ln \rho_0} e^{i \ln \rho_G(0) t}} = \Tr{\rho_0^{1+it}} D^{it}
\end{equation}
since $\rho_G(0)= I/D$ is the completely mixed state. Then, since the eigenvalues of $\rho_0$ are $\eta+(1-\eta)/D$ and $(1-\eta)/D$ which is $D-1$ fold degenerate, by evaluating the trace we get
\begin{equation}
\langle e^{it C}\rangle = D^{it} \left( (\eta+(1-\eta)/D)^{1+it}+(D-1)((1-\eta)/D)^{1+it}\right)
\end{equation}
which is Eq.~\eqref{eq. ini cohe}.
Concerning $\langle e^{iuw+itC}\rangle$ can be easily derived from the joint quasiprobability distribution of the work and coherence given in Ref.~\cite{Francica22}. By doing a symmetric choice of the parameters $q=q'=1/2$ we get Eq.~\eqref{eq. joint Cw}, from which
\begin{equation}
-i\partial_u G(0,t)= -i \partial_u \ln \langle e^{it C+ i u w} \rangle|_{u=0} = \frac{\Tr{\rho_0 e^{i \ln \rho_0 t}(H'-H)}}{\Tr{\rho_0 e^{i \ln\rho_0 t}}}
\end{equation}
and by proceeding similarly we get Eq.~\eqref{eq. deri}.

\section{Measuring the characteristic function}\label{app. meas}
The characteristic function can be measured as observed in Ref.~\cite{Francica22}. Here we note the detector can be a qubit in the initial state $\rho_D(t_i)$ with Hamiltonian $H_D = \omega \ket{e}\bra{e}$. We consider the interactions with the system described by $H_I = -\delta_e \ket{e}\bra{e} - \delta_g \ket{g}\bra{g}$ and  $H'_I = -\delta'_e \ket{e}\bra{e} - \delta'_g \ket{g}\bra{g}$, where $\ket{g}$ is the ground-state of the qubit and $\ket{e}$ is the excited state. The total system is in the initial state $\rho_D(t_i)\otimes\rho_0$ at the initial time $t_i=-t_D$, in the time interval $(-t_D,0)$ the time-evolution is generated by the total Hamiltonian $H_{tot}=H(\lambda_0)+H_D+H_I$. Then, in the time interval $(0,\tau)$ the qubit and the system do not interact and the quench is performed. Finally, in the  time interval $(\tau,\tau+t'_D)$ the time-evolution is generated by the total Hamiltonian $H'_{tot}=H(\lambda_\tau)+H_D+H'_I$. The coherence of the qubit at the final time $t_{f}=\tau+t'_D$ reads
\begin{eqnarray}
\nonumber \bra{e}\rho_D(t_{f})\ket{g}&=&\bra{e}\rho_D(t_i)\ket{g}e^{-i\omega (t_{f}-t_i)}\text{Tr}\bigg\{e^{-i(1-\delta_e)t_D H(\lambda_0)}\rho_0\\
&& \times e^{i(1-\delta_g)t_D H(\lambda_0)}U^\dagger_{\tau,0}e^{i(\delta'_e-\delta'_g)t'_D H(\lambda_\tau)}U_{\tau,0}\bigg\}
\end{eqnarray}
from which we can determine $X_q(u)$.

\end{document}